\documentclass[a4paper]{mn2e}

\usepackage{amsmath}
\usepackage{url}
\usepackage{amsfonts}
\usepackage{amsbsy}
\usepackage{graphicx}
\usepackage{subfigure}
\usepackage{verbatim}
\usepackage{amssymb}

\DeclareMathAlphabet{\mathitbf}{OML}{cmm}{b}{it}

\newcommand{\pp}{\ensuremath{\mathitbf{p}}}
\newcommand{\B}{\ensuremath{\mathitbf{B}}}
\newcommand{\kk}{\ensuremath{\mathitbf{k}}}
\renewcommand{\u}{\ensuremath{\mathitbf{u}}}
\renewcommand{\d}{\ensuremath{\partial}}

\newcommand{\ex}{\ensuremath{\mathitbf{e}_{x}}}
\newcommand{\ey}{\ensuremath{\mathitbf{e}_{y}}}
\newcommand{\ez}{\ensuremath{\mathitbf{e}_{z}}}

\newcommand{\ee}{\ensuremath{\text{e}}}
\newcommand{\ii}{\ensuremath{\text{i}}}
\newcommand{\ch}{\ensuremath{\text{ch}}}
\title[MRI channel flows in vertically-stratified models of accretion disks]{MRI channel
  flows in vertically-stratified models of accretion disks}
\author[Henrik N. Latter, Sebastien Fromang, Oliver Gressel]
{Henrik N. Latter$^{1}$\thanks{Email:
henrik.latter@lra.ens.fr}, Sebastien Fromang$^{2,3}$\thanks{Email: sebastien.fromang@cea.fr},
 Oliver Gressel$^{4}$\thanks{Email: o.gressel@qmul.ac.uk} \\ \\
$^{1}$ LERMA-LRA, \'Ecole Normale Sup\'erieure, 24 rue Lhomond, 75231 Paris Cedex 05, France\\
$^{2}$ CEA, Irfu, SAp, Centre de Saclay, 91191 Gif-sur-Yvette, France\\
$^{3}$ UMR AIM, CEA-CNRS-Univ. Paris VII, Centre de Saclay, 91191
  Gif-sur-Yvette, France\\
$^{4}$ Astronomy Unit, School of Mathematical Sciences, Queen Mary, University
  of London, Mile End Road, London, E1 4NS, UK}

\begin{document}

\maketitle

\begin{abstract}
Simulations of the magnetorotational instability (MRI) in
 `unstratified' shearing boxes exhibit powerful 
coherent flows, whereby the fluid vertically splits into countermoving planar
jets or `channels'. Channel flows correspond to certain axisymmetric linear MRI modes,
and their preponderance follows from the
remarkable fact that they are
 approximate nonlinear solutions of the MHD equations in
the limit of weak magnetic fields.
We show in this paper, analytically and with one-dimensional
numerical simulations, that this
property is also shared by certain axisymmetric MRI modes in
 \emph{vertically-stratified} shearing boxes.
These channel flows rapidly capture significant amounts of
 magnetic and kinetic energy, and thus are
 vulnerable to secondary shear instabilities.
 We examine these
 parasites in the vertically stratified context,
 and estimate the maximum amplitudes that channels attain before they are
 destroyed.
 These estimates suggest that a
dominant channel flow will usually drive the disk's magnetic
 field to thermal strengths. The
prominence of these flows and their destruction place enormous
demands on simulations, but channels in their initial stages also offer
a useful check on numerical codes. These benchmarks are especially valuable
given the increasing interest in the saturation of the stratified MRI. Lastly
 we speculate on the potential connection between `run-away' channel flows and
  outburst behaviour in protostellar and dwarf nova disks.

\end{abstract}

\begin{keywords}
accretion, accretion disks --- MHD --- instabilities --- turbulence
\end{keywords}

\section{Introduction}

Differential rotation and weak magnetic fields conspire to destabilise laminar
flow in accretion disks (and other astrophysical settings)
 via the magnetorotational instability
(MRI) (Balbus and Hawley 1991). The MHD turbulence that this linear instability sustains
can transport significant
angular momentum, and thus drive the mass accretion that is observed in many
disk systems (Hawley et al.~1995, Stone at al.~1996, Balbus and Hawley 1998, Hawley
2000).  

The characteristics of MRI turbulence have been studied extensively with local
three-dimensional
simulations in the `unstratified' shearing box\footnote[1]{Throughout the
  paper we adhere to the common (mis-) use of the term `stratified', 
  by which we mean a
  background variation in density and pressure but not necessarily in
  entropy. Normally, however, `stratified' explicitly means that the background
  entropy does vary.},
 which is a model that
 mimics a small block of gas deeply embedded in the
disk, ignorant of its large-scale vertical and radial
 structure (Balbus and Hawley 1998).
When a net vertical magnetic flux passes through such a box, the fluid
can be dominated by coherent
structures (often recurring) called channel flows. Akin to
Kolmogorov shear flow, these consist of several
 countermoving planar streams of fluid situated at different vertical levels
 (Meshalkin and Sinai 1961, Hawley et
 al.~1995, Sano and Inutsuka 2001, Sano 2007, Goodman and Xu 1994, hereafter GX94).
 Previous work has
investigated the role these structures, and their destruction,
 play in controlling MRI saturation in
the shearing box model (GX94,
 Latter et al.~2009, hereafter LLB09, Pessah and
Goodman 2009, Pessah 2009). In this paper we generalise these analyses to 
the case of a shearing box in which the \emph{vertical structure} of the disk has
been incorporated.

 The preponderance of channel behaviour in unstratified boxes
 is connected to the fact that the
channels correspond not only to linear MRI modes but also to approximate
 nonlinear solutions of the MHD equations, when in the incompressible, weak field limit (GX94). 
 In this paper, we 
show explicitly that analogous linear modes in a vertically-stratified box possess the exact same
 property --- they
 are also approximate nonlinear solutions\footnote[2]{Alternatively one could
 state that the Goodman and Xu channels are exact nonlinear solutions of the
 equations of incompressible MHD, and the solutions we present in this paper
 are exact
 nonlinear solutions of the equations of \emph{anelastic} MHD. But, since both
 classes of solution
 leave each regime so rapidly, we prefer to call them approximate
 solutions to the full equations.}.
  This property has important consequences. 
 It means that channel flows also dominate stratified
 simulations with a net vertical flux,
 growing exponentially and free of nonlinear interactions for a time
 sufficient to accumulate
 significant amounts of kinetic and magnetic energy.
 But the ultimate 
 destruction of such a large amplitude channel
  can be so violent that the computational domain is evacuated of mass, and the simulation
  compromised (Stone et al.~1996, Miller and Stone 2000). In
 contrast, behaviour elicited by other magnetic configurations (a purely
 azimuthal field, for example) is not nearly so striking, nor numerically
 problematic (Miller and Stone 2000). As a result, published simulations with a net
 vertical field are comparatively rare, even if the magnetic configuration
 itself is of astrophysical importance.

The agents responsible for the disruption of isolated channel flows are
nonaxisymmetric parasitic instabilities, which feed on their velocity and magnetic shear
(GX94, LLB09, Pessah and Goodman 2009, Pessah 2009). In general, these
parasitic modes are ineffectual until large channel amplitudes are attained because
of the combined action of the background shear and the exponential growth of the
channel itself. In this paper, we compute directly some of the ideal MHD parasitic modes that might be
present in stratified boxes in a convenient intermediate limit. Using these
results, we estimate the minimum channel amplitude that permits a
successful parasitic attack.
In a later article a general survey with three-dimensional simulations will be
presented.

The paper is organised as follows. First, in Section 2 we present linear MRI
channel modes in a vertically-stratified box
 and show that they are also nonlinear solutions in the limit of
weak magnetic fields. These are directly computed numerically and then checked with
one-dimensional stratified simulations in Section 3. Following that, in
Section 4, the ideal
MHD parasitic instabilities which could disrupt these channels are
characterised and discussed.
 Our conclusions are drawn in Section 5, where we also briefly discuss the potential
 role of these solutions in real accretion disks.

\section{MRI channel modes as nonlinear solutions}

We study the MRI in a `quasi-global' setting, in
which the vertical structure of the disk is included but the radial
structure is neglected (except, of course, for the Keplerian shear). 
In the presence of weak magnetic fields, the essence of the MRI is local,
and as such is relatively impervious to the boundary conditions and larger-scale
properties of the disk. Indeed, the nature of the turbulence induced by the
MRI has been studied profitably in the local setting of the shearing box for
over a decade. However, the inclusion of vertical structure can help us
 address important questions such as the
relationship between the MRI, on the one hand, and on the other: magnetic buoyancy, 
depth-dependent ionisation,
 realistic radiation effects, and the formation of a disk-corona and
 large-scale magnetic fields
 (see for example, Brandenburg et al.~1995, Stone et al.~1996,
  Johansen and Levin 2008, Davis
   et al.~2010, Shi et al.~2010, Sano and Miyama 1999, Stone and Fleming 2002,
   Miller and Stone 2000, Hirose et al.~2006, Blackman and Pessah 2009).

The first linear axisymmetric stability analysis of the ideal MRI in a 
vertically-stratified local model
was undertaken by Gammie and Balbus (1994) 
(see also the full global treatment of Papaloizou and Szuszkiewicz 1992).
 They addressed different equilibrium magnetic
field configurations in addition to various vertical 
boundary conditions. Generally, the MRI
eigenmodes must be calculated numerically, though some analytic progress can
be made when the field is purely vertical
and if it is assumed that the gas extends over all $z$ (the `infinite disk'
model). These results were generalised in a more realistic geometry by Ogilvie
(1998). 
 A number of papers since have numerically
computed the linear eigenmodes, usually in the context of the non-ideal MHD
prevalent in protostellar disks (see, for example, Sano and Miyama
1999, Salmeron and Wardle 2003, 2005, 2008). More recently, Liverts and Mond
(2009) returned to the isothermal ideal MHD infinite disk and derived analytic
approximations to the linear modes with matched asymptotics in the limit of
large plasma beta.

In this paper we confine ourselves to the simple case of ideal MHD and 
an infinite
disk threaded with a vertical
magnetic field. Though
 the axisymmetric linear MRI modes in this setting have been well-established, what has
 perhaps not been fully appreciated is that for large midplane $\beta$ a
 subset of these linear
 modes (the channel flows) are approximate nonlinear solutions as well, a fact independent of the
 particulars of the vertical stratification. 
This property of  stratified MRI channel flows
 has been suggested by some
simulations (Miller and Stone 2000) and by `hybrid' MRI-wind solutions
(Ogilvie and Livio 2000, Salmeron et al.~2007). 
Here we demonstrate the result explicitly.

\subsection{Governing equations}

We study a disk of perfectly conducting fluid orbiting a massive central
point with a frequency $\Omega(r)$ (where $r$ is cylindrical radius).
Consequently, we employ the equations of ideal MHD and
frame these in the geometry of the stratified shearing sheet (Goldreich and
Lynden-Bell 1965, Gammie and Balbus 1994).
 This is the standard local approximation that describes a `block of disk'
 centred on the midplane and at a radius $r_0$, and moving along
 the circular orbit that radius
 prescribes. The Cartesian coordinates $x$ and $y$ of the block correspond to the
 radial and azimuthal directions respectively. The differential
 rotation is accounted for by the 
Coriolis force and a background linear shear, $\u=
 -2\, A_0\,x\,\ey$, where 
$$ A_0 = -\frac{r_0}{2}\left(\frac{d\Omega}{dr}\right)_0. $$ 
A Keplerian disk yields $A_0=3/4$ and throughout this paper a
Keplerian rotation law will always be assumed. In addition, to account for the vertical gravity of
the central object, we include the force $-\Omega^2_0\,z\,\ez$ in the momentum
equation (with
$\Omega_0=\Omega(r_0)$), which is simply the
first term in an expansion of the gravitational potential in $z$. The approximation breaks down for large $z$, therefore
we have
assumed that the disk is relatively thin. From now the subscript `0'
will be dropped.

The governing equations of the problem are
\begin{align} \label{gerho}
&\d_t \rho = -\nabla\cdot(\rho\,\u), \\
&\d_t \u + \u\cdot\nabla\u =   - \nabla \Phi -2\Omega \ez\times \u \notag  \\
& \hskip3cm -\frac{1}{\rho}\nabla \left( P+ \frac{B^2}{8\pi}\right) 
 + \frac{\B\cdot\nabla\B}{4\pi\,\rho}, \label{geu}\\
& \d_t\B = \nabla\times(\u\times\B), \label{geB}
\end{align}
where the tidal potential is defined through
\begin{equation}
 \Phi = -\tfrac{3}{2}\Omega^2 x^2 + \tfrac{1}{2}\Omega^2 z^2.
\end{equation}
To this set we must add the equation of state and possibly the internal energy
equation. We must also ensure that the magnetic field $\B$ is solenoidal.

\subsection{Equilibrium and solution ansatz}

Let us suppose the disk is pierced by a weak vertical magnetic field. At some large
height the magnetic field lines will most likely arc back and reenter the disk
at a different radius, or perhaps plunge into the central object itself, but
we neglect this larger structure, and consequently ignore any local force that a
magnetic field line may exert due to the influence of its remote
`footpoint'. We also ignore the possibility that the field lines are embedded
in a hot magnetic halo. (Both cases are treated in Gammie and
Balbus 1994.) 
 The governing equations then
 admit the following equilibrium state:
\begin{align*}
 &\u_0 = -(3/2)\Omega\,x\,\ey, 
 &\rho_0 =\rho_{00}\,\,h\left(\frac{z}{H}\right), \\
 & P_0 =P_0\left( \frac{z}{H}\right), 
 &\B_0 =B_0\, \ez.
\end{align*}
In the above, $h$ is a dimensionless symmetric function and $H$ is the characteristic lengthscale of
the disk's vertical structure. The dimensional $\rho_{00}$ and $B_0$ are
 constants. In addition, there may be variations in temperature and entropy.  
 The two functions $h$ and $P_0$ can be determined from the steady force and
 energy balances once the equation of state is specified, but their exact
 forms are unimportant in what follows. We
 require only that the stratification is convectively stable. 

Consider a perturbation to the basic
 state, dependent on only $z$ and $t$, 
and taking the form of a planar channel flow:
\begin{align} \label{F}
\u^{\ch} &= b\,u_0\,F\left( \frac{z}{H}\right)\,\ee^{st}\,[ \ex\, \cos\theta +
\ey\,\sin\theta], \\
\B^{\ch} &= b\,B_0\, G\left( \frac{z}{H}\right)\,\ee^{st}\,[ \ex\,\sin\theta -
\ey\,\cos\theta], \label{G}
\end{align}
where $F$ and $G$ are dimensionless functions, $s$ is a growth rate,
$\theta$ is a constant orientation angle, and $u_0$ and $B_0$
constants, the latter introduced earlier in the equilibrium solution. Note
also the dimensionless constant $b$, which measures the relative
amplitude of the channel to its background at $t=0$. These perturbations
constitute the basic MRI.

It is easy to check that the profiles Eqs \eqref{F}-\eqref{G} 
possess the following attractive properties:
\begin{align*}
&\u^\ch\cdot\nabla\u^\ch=\B^\ch\cdot\nabla\B^\ch=0, \\
& \u^\ch\cdot\nabla \B^\ch= \B^\ch\cdot\nabla\u^\ch=0.
\end{align*}
These cancellations mean that all the nonlinear terms in the induction
equation and most of the nonlinearities in the momentum equation disappear.

Via magnetic pressure, the MRI channel flow  
drives perturbations
in the thermodynamic variables, $\rho^{\ch}$, $P^{\ch}$, etc,
but these will remain small relative to the background equilibrium for as long as the
 magnetic field (the driving) is weak. We assume that the channel's magnetic
 field is indeed weak, at least initially,
 and consequently \emph{linearise} the governing equations 
in the thermodynamic perturbations, more specifically $\rho^\ch$.
 The continuity equation
 becomes
\begin{equation}
\nabla\cdot(h\,\u^{\ch})=0,
\end{equation}
to leading order, which is satisfied by construction, and the thermodynamic
perturbations drop out entirely from the $x$ and $y$
components of the momentum equation, i.e.\ $1/\rho$ is approximated by
$1/\rho_0$ in the magnetic tension terms and we obtain
\begin{align}
\d_t\u^\ch = -2\Omega\ez\times\u^\ch + \frac{B_0}{4\pi\,\rho_0(z)}\frac{\d
  \B^\ch}{\d z}.
\end{align}
(Now all nonlinearities have vanished from these equations.)
As a consequence, the $z$ component of the momentum equation,
\begin{equation} \label{P}
\frac{d}{dz}\left[P^{\ch}+\frac{(B^{\ch})^2}{8\pi}\right] + \Omega^2\, z\, \rho^{\ch}=0,
\end{equation}
 the equation of state, and the entropy equation are \emph{subordinate} to
the MRI dynamics. This set of `passive' equations are acted upon by the MRI equations,
 via the channels' magnetic
pressure, but not vice-versa. 
Their sole job is to
determine the small thermodynamic fluctuations driven by the
channels and we are not obliged to solve them explicitly. Our only concern is
for the magnitudes of these fluctuations, which must remain small or else the
approximation breaks down.
From \eqref{P} it is clear that these amplitudes,
 relative to their background state,
are of order
 $\epsilon = b^2 e^{2st}/\beta$, where 
\begin{equation} \label{beta}
\beta= \frac{2H^2\,\Omega^2}{ B_0^2/(4\pi\,\rho_{00})},
\end{equation} 
is the (midplane) plasma beta associated with the background vertical
field. Our linearisation in $\rho^\ch$ 
is only valid for as long as $\epsilon$ remains small.

This
formalism is actually the anelastic approximation, modified so as to
accomodate the
MRI (Gough 1969, Gilman and Glatzmaier 1981, Barranco and
Marcus 2005). Notice that the
incompressible limit employed by GX94 is a special case,
obtainable if we further assume the channel flows vary
 on scales much smaller than $H$.

We conclude that the channel flow ansatz introduced in Eqs \eqref{F}-\eqref{G} is not only
a linear solution but also an approximate \emph{nonlinear solution} to the
governing equations
in the regime of weak magnetic field (or, equivalently, small thermodynamic
fluctuations).
 Because the approximation introduces errors of order $\epsilon^2$, we expect 
the ansatz
to fail when $\epsilon\sim 1$.
Owing to the channels'
 exponential growth, this regime is achieved
 rapidly unless other agents intervene (such as parasitic
 instabilities). The time at which the approximation fails can be simply estimated
 by
\begin{equation}
t \sim \frac{1}{2s}\ln (\beta/b^2).
\end{equation}
With a typical MRI growth rate and $b=0.1$ and $\beta=10^4$, this time
corresponds to only 1.5 orbits. But it also means
 that, in principle, a single MRI stratified channel
can generate equipartition strength fields from very weak fields in astonishingly short
times, just as its unstratified counterparts can (GX94, Hawley et al.~1995,
LLB09).  
This is in fact witnessed in the stratified simulations of Miller and
Stone (2000).

\subsection{The eigenproblem for channel flows}

We now supply the remaining details of the ansatz by deriving equations for
$F$ and $G$ and a dispersion relation for $s$.
 The solution \eqref{F}--\eqref{G} is substituted into the governing equations so that
$ \u=\u_0+\u^\ch$ and $\B=\B_0+\B^\ch$. After linearising in $\rho^\ch$,
 the four equations governing the MRI
channel flow are
the $x$ and $y$ components of the momentum and induction equations. These are
\begin{align}\label{e1}
&u_0(s\,\cos\theta-2\Omega\,\sin\theta)\,F=
(v_A)^2\,\sin\theta\left(\frac{1}{h}\frac{dG}{dz}\right),\\
& u_0(s\sin\theta+\frac{1}{2}\Omega\cos\theta)\,F=
-(v_A)^2\cos\theta\left(\frac{1}{h}\frac{dG}{dz}\right),\\
& s\,\sin\theta\,G = u_0\cos\theta\,\frac{dF}{dz}, \\
& (-s\,\cos\theta + \frac{3}{2}\,\sin\theta)\,G = u_0\sin\theta\,\frac{dF}{dz},\label{e2}
\end{align}
where $v_A= B_0/\sqrt{4\pi\rho_{00}}$ is the equilibrium 
Alfv\'en speed at the midplane. The four
functional equations can be satisfied if
\begin{align} \label{rels}
F = - \frac{1}{Kh}\frac{dG}{dz}, \qquad G =\frac{1}{K} \frac{dF}{dz},
\end{align}
which together return the linear second-order ordinary differential equation:
\begin{equation} \label{master}
\frac{d^2F}{dz^2} + K^2\,h\,F=0,
\end{equation}
where the `vertical wavenumber' $K$, a constant, is yet to be determined.
Note that Equation \eqref{master} was also derived by Gammie and Balbus
 (1994) (cf.\ their Eq.~(23)).

Equation \eqref{master} with suitable boundary conditions
 describes a 1D eigenvalue problem for the vertical MRI modes
with $K$ as eigenvalue. Because $h$ is always positive the equation is in
classical Sturm-Liouville form and so we are assured of a discrete set of
real eigenvalues $\{K_n\}$ and eigenfunctions $\{F_n\}$. Moreover, the
eigenfunctions are orthogonal under integration with weight $h$ (Arfken
1970).  The solutions possess
the symmetry $(K_n,\,F_n,\,G_n)\to(-K_n,\,F_n,\,-G_n)$. 
Notice that the lowest order solution is trivial, $K_0=0$ and $F$ a
constant, but this does not correspond to the MRI. 

 The boundary condition comes from the requirement that
 the Alfven speed of the perturbation $(v_A^\ch)^2 \propto G^2/h$ must never
diverge. Thus $G$ (and hence $dF/dz$) must go to zero at the
disk boundaries.
Solutions to Eq.~\eqref{master} are computed in the following 
subsection for various equilibria.

Returning to Eqs \eqref{e1}-\eqref{e2}, the $z$ functions factor out and we
are left with the unstratified incompresible, but discrete,
 dispersion relation of the MRI. Eliminating
$\theta$ and $u_0$ we get for a given $n$ mode
\begin{equation} \label{disp}
 s^4 + (\Omega^2+2 v_A^2 K_n^2) s^2 + v_A^2 K_n^2(v_A^2 K_n^2-3\Omega^2)=0.
\end{equation}
If time and space are scaled by $\Omega^{-1}$ and $H$, respectively, the dispersion relation only
depends on the two dimensionless quantities $\beta$ and $n$. Evidently, the basic physics
of the MRI is shared at each separate $z$, and the instability is not dramatically altered
by the stratification. Note that Eq.~\eqref{disp} was
recovered in Gammie and Balbus (1994) and Liverts and Mond (2009). In the
latter it appears in their Eq.~(19), if one sets $K_n= \pi(n+1/2)/\Phi(-z_0)$ 
in their notation.
 To complete the
description we need $\theta$ and $u_0$, which can be
computed from
\begin{align}\label{theta}
\sin^2\theta &= -\frac{1}{6}\left(1-\sqrt{1+32 \frac{K_n^2 H^2}{\beta}} \right), \\
u_0 &= \frac{3\,\Omega}{2 K_n }\sin^2\theta. \label{u0}
\end{align}
The quantity $u_0/(H\Omega)$ may be regarded as the Mach number of the channel.
Exactly as in the unstratified MRI, the fastest growing mode possesses the
orientation angle nearest $\pi/4$, and the flow speed is always sub-Alfv\'enic
(GX94, LLB09). Finally, Eq.~\eqref{theta} yields a critical $\beta$ below which there
can be no MRI at all: we find the critical value to be $\beta_c=(32/15)(K_1
H)^2$. Because the longest mode ($n=1$)
will generally possess a wavenumber $K_1\gtrsim 1/H$, we obtain $\beta_c\gtrsim 2$.
This is the well-known linear result that instability is
suppressed when the magnetic field (in fact, the magnetic torsional stress)
is too strong (Balbus and Hawley 1991). 
Of course, this low $\beta$ regime is not relevant for our
 large $\beta$ nonlinear solutions.

\subsection{Examples}

\subsubsection{Unstratified disk}

The unstratified theory (Balbus and Hawley 1991, GX94) 
is recovered by setting $h=1$ and sending the vertical boundary conditions to
positive and negative infinity.
 Equation
\eqref{master} is then the harmonic oscillator equation, and
\begin{equation}\label{unstr}
 F= \sin(K z).
\end{equation}
Because the boundary conditions have been dropped, 
the vertical wavenumber $K$ is a free parameter, 
and because $H$ has disappeared from the problem the plasma beta has
as well. Equation \eqref{unstr} also corresponds to the `local' (WKBJ) solution 
of \eqref{master}, which assumes that the eigenfunction varies on scales much
shorter than $H$.

\subsubsection{Isothermal disk}

The isothermal ideal gas equation of state is $P=c_s^2 \rho$, where $c_s$ is
the constant sound speed. This yields a Gaussian equilibrium density stratification:
 $h=\exp[-z^2/(2H^2)]$, with $H=c_s/\Omega$. In this case the eigenvalue equation cannot be solved
analytically. But it is a straightforward task to compute the eigenfunctions
using a pseudo-spectral method. We solve for $G$ rather than $F$ as it decays
as $|z|\to\infty$ and so we can employ a basis of Whitakker cardinal
functions (Boyd 2001, also see Section 4.4)\footnote[1]{The numerical script
  that solves the isothermal eigenproblem is available
  on email request}. MRI channel eigenfunctions are plotted in Fig.~1 for
$n=1$ to $4$ and $n=10$. 

As is clear, the MRI flow resemble a vertical sequence of planar `channels' or
`jets'. The centre of each jet corresponds to a magnetic null surface. This
is a general result which follows from the second equation
in \eqref{rels}: a maximum of $F$ is a zero of $G$. The fluid velocities ($F$)
go to
 constant values as $z\to \infty$, while the magnetic fields ($G$) go to zero,
 in accordance with the boundary conditions.
 When $n$ is an even integer there exist $n-1$ 
jets, one centred at $z=0$, and symmetric pairs on each side at $\pm z_i$ for
$i=1,\dots (n-2)/2$. The $z_i$ are the zeros of the function $G_n$. When $n$ is
odd there are $n-1$ jets occuring in antisymmetric pairs, and these are superimposed
on a large-scale shear flow. In Fig.~1 this pattern can be observed;
 the somewhat special 
$n=1$ case corresponds to a simple shear flow (no jets), the $n=2$ case
 corresponds to a single jet centred at the midplane, the $n=3$ case
 corresponds to a double jet superimposed on a shear flow, the $n=4$ case to a
 triple jet, and so on. Finally, note that because $K_1=1.158$, we have $\beta_c=
 2.86$.

\begin{figure}
\scalebox{0.55}{\includegraphics{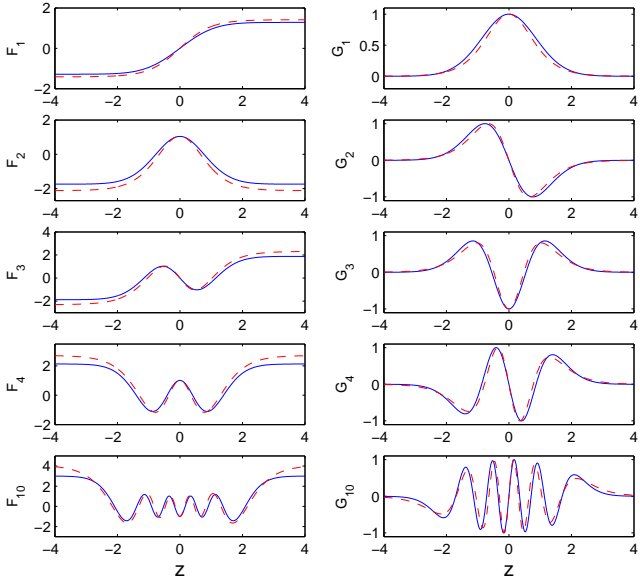}}
\caption{Eigenfunctions $F_n$ and $G_n$ of order $n=1,2,3,4$ and $10$ computed numerically for the isothermal model of
  Section 2.4.2. The functions are plotted with solid lines and are normalised
  so that the maximum of $|G_n|$ is 1.  The corresponding
  wavenumbers are $K_nH=1.1584$, $2.0796$, $2.9829$, $3.8798$ and
  $K_{10}H =9.2239$.
 The dashed lines are the corresponding
  approximations of Section 2.4.3, using the Legendre polynomials. The corresponding
  eigenvalues are  $1.4142$, $2.4495$, $3.4641$, $4.4641$, $10.4881$. }
\end{figure}

\subsubsection{Approximate isothermal model}

 A rough approximation to the isothermal
stratification is the profile $h=\text{sech}^2(z/H)$ 
(which also describes a purely self-gravitating layer --- Gammie
and Balbus 1994). The advantage of this choice is that it 
furnishes us with an analytic solution to Eq.\
\eqref{master}. After some straightforward
analysis one obtains
\begin{align} \label{Legendrefuck}
 F_n = \text{P}_n[ \text{tanh}(z/H)], \quad K_n = \frac{1}{H}\sqrt{n^2+n},
\end{align}
where $\text{P}_n$ is a Legendre polynomial of order $n$ (Arfken 1970). 
These approximations are plotted in Fig.~1 as dashed lines alongside the
numerical isothermal eigenfunctions. As is clear, the Legendre polynomials do
an adequate job at replicating the isothermal channels, with the most
noticeable discrepancies at large $|z|$.

\subsubsection{Polytropic disk}

A polytropic disk can be described by $P=C\,\rho^{1+1/m}$ where $m$ is the
polytropic index and $C$ is a dimensional constant. Such a disk will settle into an equilibrium density stratification of the form
$h= (1-z^2/H^2)^m$, where $H$ is a combination of $m$, $C$, $\Omega$, and
$\rho_0$, and denotes the actual disk surface --- there exists only vacuum where
$|z|>H$. For general $m$, the boundary value problem \eqref{master} must be
solved numerically, though an analytical solution in terms of parabolic
cylinder functions is available for the special case of $m=1$. 
In Fig.~2 numerically computed solutions are presented
when $m=3/2$, which corresponds to an ideal gas with an adiabatic index of
$5/3$. These are produced by a simple shooting method.
 We plot the first four nontrivial eigenfunctions as well as the $n=10$ solution.
The critical plasma beta, below which there exists no unstable MRI,
 is $\beta_c=13.8$ when $m=3/2$.

\begin{figure}
\scalebox{0.6}{\includegraphics{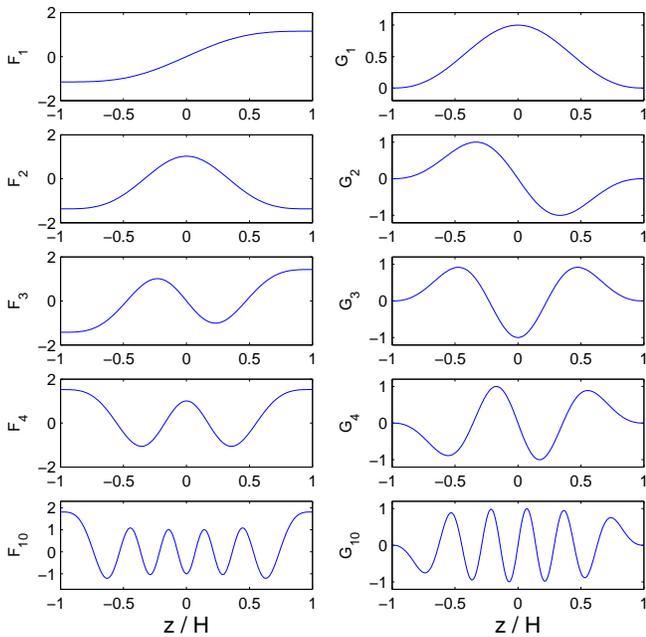}}
\caption{Eigenfunctions $F_n$ and $G_n$ of order $n=1,2,3$ and $10$ computed
  numerically for 
the polytropic model of
  Section 2.4.4 with $m=3/2$. The corresponding
  wavenumbers are $K_nH=2.541$, $4.762$, $6.963$, and $22.29$
  respectively. The profiles resemble those in Fig.~1 but are confined between
$z=\pm H$.}
\end{figure}

\subsection{Discussion}

The `nonlinear property' of the channel modes permits the accumulation of
 significant
 kinetic and magnetic
energies in the first few orbits
of the MRI evolution. At large midplane $\beta$, small initial
 conditions seed a suite of fast-growing channel modes, which
 continue growing independently even once they reach the magnitude of the background
 vertical field $\B_0$. Because channels modes do not nonlinearly interact
 amongst themselves there may be no
 check on their growth until equipartition amplitudes.

There are perhaps two ways to avoid this outcome. First, parasitic modes, feeding off
the MRI channels' strong velocity shear may destroy the channel prematurely
(GX94, LLB09, Pessah and Goodman 2009, Pessah 2009). We explore this in more detail in Section
4, but both unstratified and stratified simulations indicate that parasitic
modes generally cannot halt the channels' march to equipartition in the early
stages of a run (Hawley et
al.~1995, Miller and Stone 2000, LLB09). Second, channel flows may interact
nonlinearly with slower-growing radially
dependent MRI modes or, in particular, transiently growing non-axisymmetric modes (Balbus and
Hawley 1992, Hawley et al.~1995). In principle, if the amplitude
of the initial conditions is set equal to or above that of the background $B_z$,
a non-axisymmetric MRI mode (shearing wave) could destructively interfere with a channel flow,
before it dominates everything else. Indeed, seeding large-amplitude noise is a common
technique in unstratified mean vertical flux simulations, as it averts the initial channel `spike' (Hawley
et al.~1995, LLB09). This technique should also serve in stratified simulations
at intermediate and lower
$\beta$. In very large $\beta$ runs (as in Suzuki and Inutsuka 2009), the
seeded noise is usually nonlinear from the very start.

If a channel reaches equipartition, magnetic pressure will be sufficient
to alter the channel structure, squeezing mass onto the horizontal sheets
where the magnetic field is zero. Some of this
behaviour is explored in the following section with one-dimensional numerical
simulations. The stability of these `extreme channel' were studied in the
unstratified context in LLB09, which showed they were
undone by compressible and resistive parasitic modes.

\section{Isothermal 1D simulations}

In this section we present a set of 1D numerical simulations using two
different finite-volume 2nd order Godunov MHD codes. The purpose of these simulations is twofold.
First, they will serve as a validation of the ansatz in Section 2, while
conversely providing a useful numerical check on these codes. 
Their second aim is to determine how
far into the nonlinear regime the linear channels can be followed
before magnetic pressure effects become important. Obviously, an important
limitation of the simulations is that they are 1D. As
such, they cannot capture the parasitic modes
considered in Section 4, which should be studied
with 3D numerical simulations.

\subsection{Numerical setup}

We used the MHD codes RAMSES (Teyssier 2002, Fromang et al.~2006) and
NIRVANA (Ziegler 2004) to solve Eqs \eqref{gerho}-\eqref{geB}
along the $z$ direction only. An isothermal equation of state
is adopted and, accordingly, the density profile at the start of the
simulation is set to a Gaussian. A pure vertical magnetic field is added to
this hydrostatic equilibrium, whose strength is governed by the midplane
$\beta$. The computational
domain extends in all cases from $z=-5H$ to $z=+5H$, which we found
was the minimum size needed for accurate results.
 The vertical boundary conditions are as follows: the density
is extrapolated in the ghost zones in order to satisfy the vertical
hydrostatic equilibrium, outflow boundary conditions are applied on
the velocities, and zero-gradient boundary conditions are applied to
the tangential component of the magnetic field (in 1D, the vertical
field is constant in space and time). At $t=0$, we either seed the
isothermal eigenmodes calculated in
Section 2.4.2 or their approximation in terms of
Legendre polynomials, outlined in Section 2.4.3. In
both cases, their amplitudes are such that the maximum value of the
horizontal magnetic field is a fraction $b$ of the vertical
field $B_0$.

\subsection{The case $\beta=10^2$}

\begin{table}
\begin{center}
\begin{tabular}{@{}cccc}\hline\hline
Mode & $\sigma_\text{th}/\Omega$ & $\sigma_\text{num}/\Omega$  & $\sigma_\text{num}/\Omega$ \\
number &  & {\it RAMSES} & {\it NIRVANA} \\
\hline
\hline
\multicolumn{4}{c}{$\beta=100$} \\
\hline
n=1 & $0.2664$ & $0.2662 \pm 0.0001$ & $0.2674 \pm 0.0009$\\
n=2 & $0.4307$ & $0.4302 \pm 0.0001$ & $0.4276 \pm 0.0001$\\
n=3 & $0.5502$ & $0.5495 \pm 0.0002$ & $0.5471 \pm 0.0002$  \\
n=4 & $0.6363$ & $0.6353 \pm 0.0003$ & $0.6330 \pm 0.0002$ \\
n=5 & $0.6957$ & $0.6944 \pm 0.0003$ & $0.6920 \pm 0.0002$ \\
n=6 & $0.7326$ & $0.7307 \pm 0.0003$ & $0.7290 \pm 0.0002$ \\
n=7 & $0.7489$ & $0.7463 \pm 0.0003$ & $0.7458 \pm 0.0003$ \\
n=8 & $0.7455$ & $0.7414 \pm 0.0004$ & $0.7427 \pm 0.0004$ \\
n=9 & $0.7214$ & $0.7153 \pm 0.0004$ & $0.7148 \pm 0.0005$ \\
n=10 & $0.6744$ & $0.6653 \pm 0.0003$ & $0.6618 \pm 0.0003$ \\
\hline
\multicolumn{4}{c}{$\beta=1000$} \\
\hline
n=20 & $0.7343$ & $0.7308 \pm 0.0008$ & $0.7174\pm 0.0005$ \\
n=30 & $0.7133$ & $0.6921 \pm 0.0038$ & $0.6989\pm 0.0012$ \\
\hline
\hline
\multicolumn{4}{c}{$\beta=100$--Legendre Approximation} \\
\hline
n=4 & $0.6363$ & $0.6357 \pm 0.0041$ & $0.6374 \pm 0.0041$\\
n=7 & $0.7489$ & $0.7413 \pm 0.0033$ & $0.7435 \pm 0.0012$ \\
n=10 & $0.6744$ & $0.6677 \pm 0.0089$ & $0.6737 \pm 0.0064$ \\
\hline
\hline
\end{tabular}
\caption{Growth rates of the normal modes of order n ({\it first
column}) obtained with the codes RAMSES ({\it third column}) and
NIRVANA ({\it fourth column}). The second colum display the theoretical
growth rates for the purpose of comparison. The data under the heading `Legendre Approximation'
were obtained with runs in which the initial condition was an approximate isothermal
eigenfunction (cf.\ Section 2.4.3); all other runs were seeded with the
correct isothermal eigenfunction (computed in\ Section 2.4.2).}
\label{growth_rates_eigen}
\end{center}
\end{table}

\begin{figure}
\begin{center}
\includegraphics[scale=0.45]{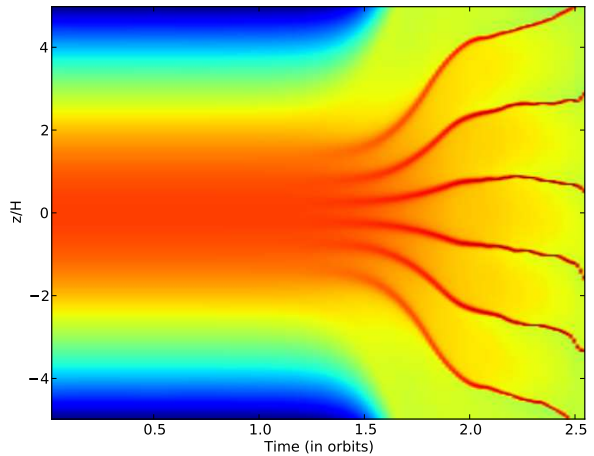}
\caption{Spacetime diagram of the gas density in a simulation
performed with RAMSES for $\beta=100$ in which a pure $n=7$ eigenmode is seeded at
time $t=0$. Density spikes develop between regions of strong magnetic field,
which then lift up the thin veins of matter through magnetic
buoyancy.}
\label{spacetime_n7}
\end{center}
\end{figure}

We first consider the case $\beta=10^2$. Admittedly, this is a choice that 
does not best demonstrate the nonlinear property of the
channel flows: the channels can grow only a few times $B_0$ before hitting
equipartition. However, lower $\beta$ runs facilitate a clean
comparison of the growth rates and eigenfunctions, because the various modes
are fewer and relatively well-spaced.

We set $b=0.01$ and seed as initial conditions the isothermal eigenfunctions
 of Section 2.4.2
 with $n$ values between $1$ and $10$. For this
range of parameters, we find a resolution of $N_z=256$ gridpoints to be
 mandatory in order to properly reproduce the mode morphology.
 This amounts to about $25$ cells per scaleheight. For each
$n$, we measure the growth rates with both
RAMSES and NIRVANA. They are reported in
Table~\ref{growth_rates_eigen}, in the third and fourth
columns respectively, while the second column gives the expected theoretical growth
rate computed using Eqs \eqref{master} and \eqref{disp} with the isothermal model.
 For all $n$, the agreement between
the numerical results and the analytical prediction is excellent,
cross-validating both results. It should be noted, though, that it is
difficult to follow the $n=1$ mode for longer than $1.5$ orbits (at
which point the eigenmode has grown by only about an order of
magnitude). Because $\sigma_\text{num}$ is so small for this mode,
 any
numerical error tends to seed higher order rapidly-growing
eigenmodes that quickly overcome the seeded $n=1$
mode.

\begin{figure*}
\begin{center}
\includegraphics[scale=0.4]{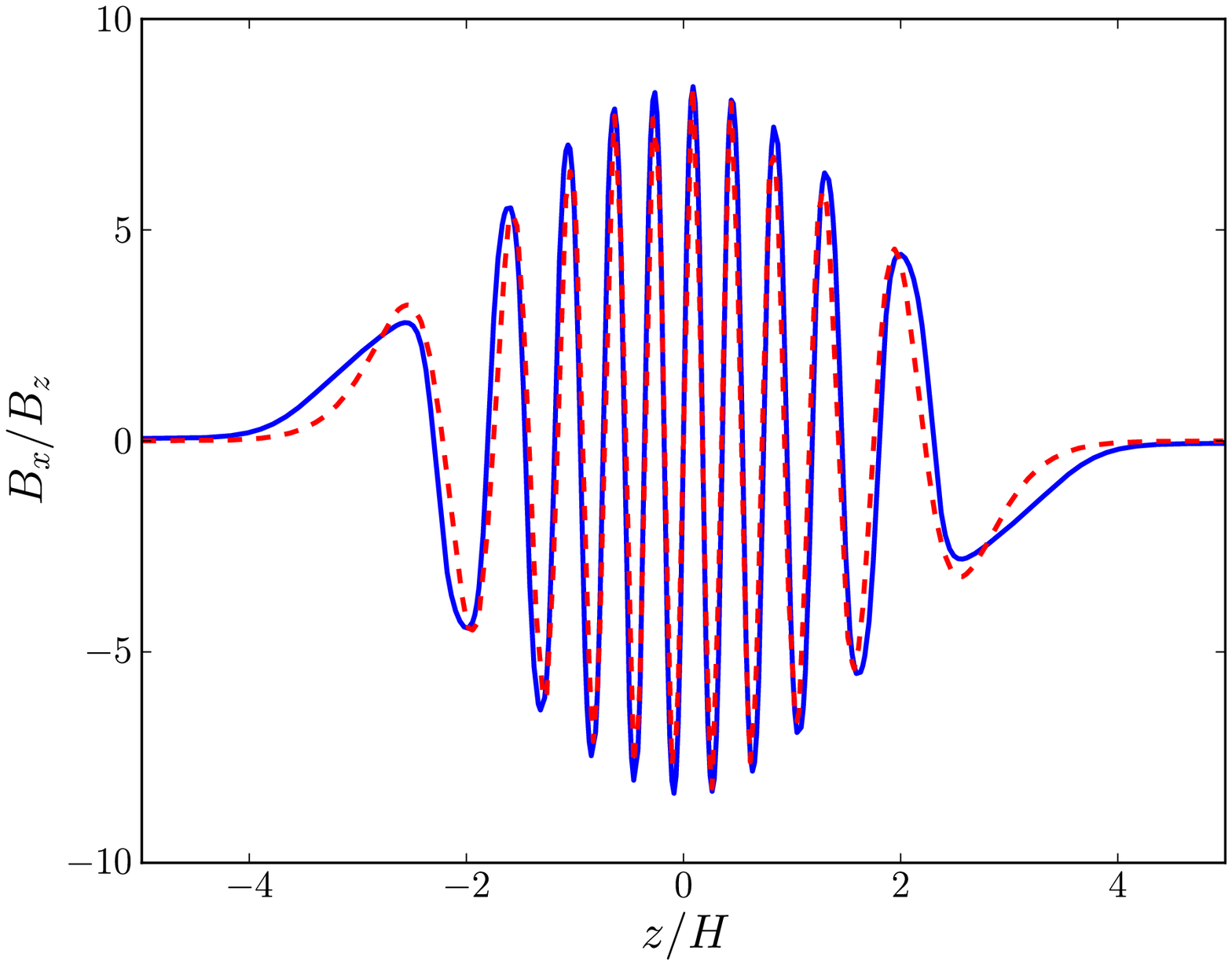}
\includegraphics[scale=0.4]{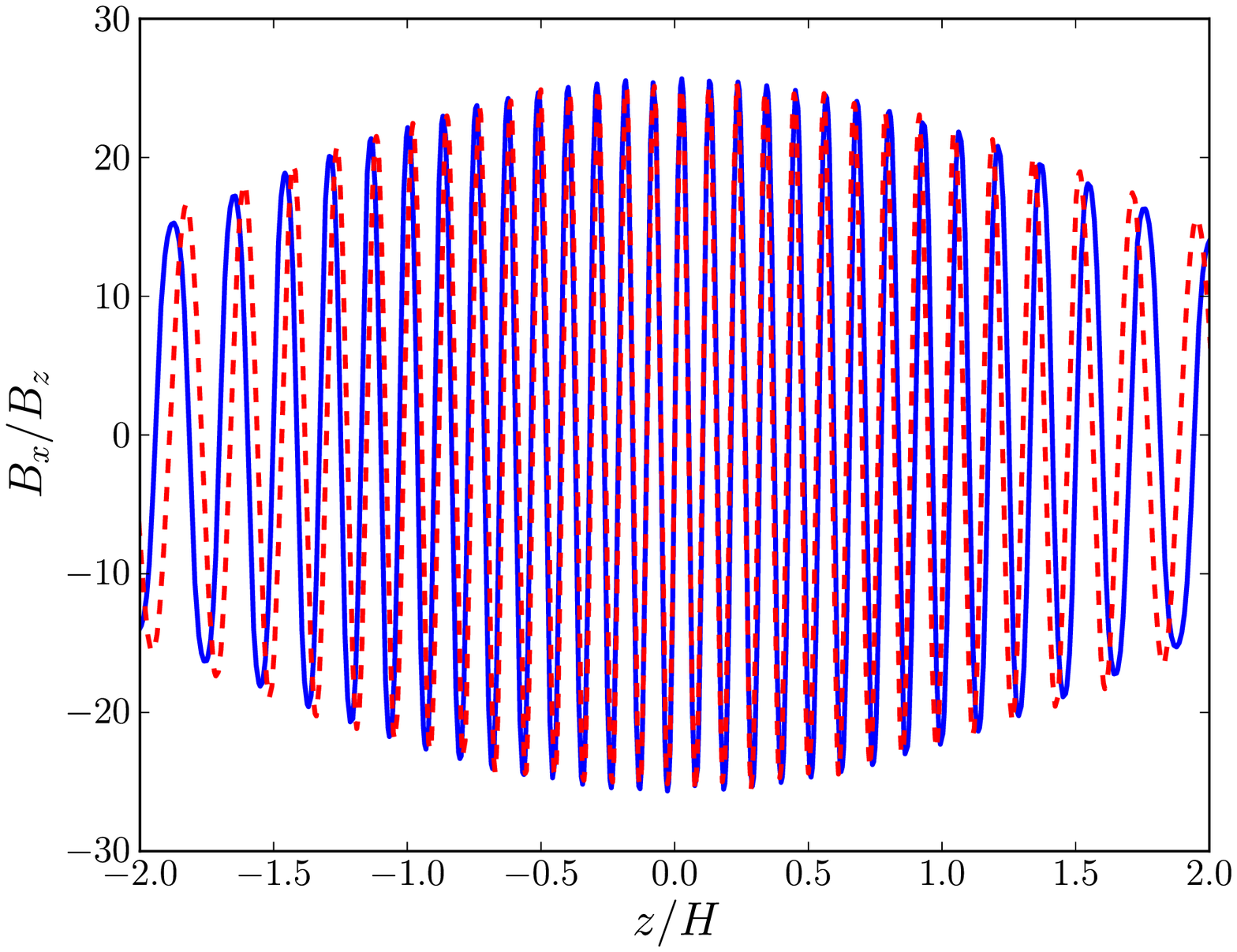}
\caption{Left panel: Vertical profile of $B_x/B_0$ ({\it solid line}) at time $t=1.1$
orbits for a simulation performed with RAMSES in which an exact normal
mode of order $n=20$ is seeded initially with $\beta=10^3$. The dashed line 
shows the vertical profile of $B_x$ at $t=0$, rescaled in order to
match the amplitude of the field at $t=1.1$. Right
panel: as in the first panel, but with $\beta=10^4$, a seeded $n=68$ mode, and
after $t=1.3$. The domain has been truncated to better display the fine-scale
structure.}
\label{bx_eigen_beta1000}
\end{center}
\end{figure*}

Finally, we comment briefly on the nonlinear evolution of the
eigenmodes. A spacetime diagram of the density field $\rho$
 for the seeding of a $n=7$ mode is
presented in Fig.~\ref{spacetime_n7}. No features are apparent when $t<1.3$
orbits, corresponding
 to the weak-field, anelastic
 phase of the evolution. At near $t=1.3$ orbits
 the amplitude of
the channel magnetic field is already a few times that
of the vertical field $B_0$, yielding 
a midplane $\beta$ of order 10. At later times, or rather smaller $\beta$,
 magnetic pressure
fluctuations compress the density into thin layers that are then lifted
up by magnetic buoyancy. This is similar to the results of past 3D
numerical simulations of the MRI in the shearing box in the presence
of a vertical flux (Miller and Stone 2000). 

\subsection{The cases $\beta=10^3$ and $\beta=10^4$}

We also perform 1D simulations with $\beta=10^3$ and $\beta=10^4$, both with
$b=0.1$. These larger $\beta$ values offer
 more opportunity for the channel flows to confirm that
they are indeed good approximate nonlinear solutions.

 When $\beta=10^3$, the fastest growing eigenmodes are those
 with $n$ values between $n=20$ and $n=30$. To adequately describe these modes,
 a resolution of
$N_z=512$ is required.  This corresponds to more than
$50$ cells per scaleheight. A lower resolution is unable to represent the high
spatial frequencies of the profiles. Despite these large spatial resolution, we
are unable to follow the growth of low order modes. This is
again because their growth rates are so comparatively small:  
numerical errors seed
higher order modes that grow faster and quickly distort their low order
precursors.

\begin{figure}
\begin{center}
\includegraphics[scale=0.45]{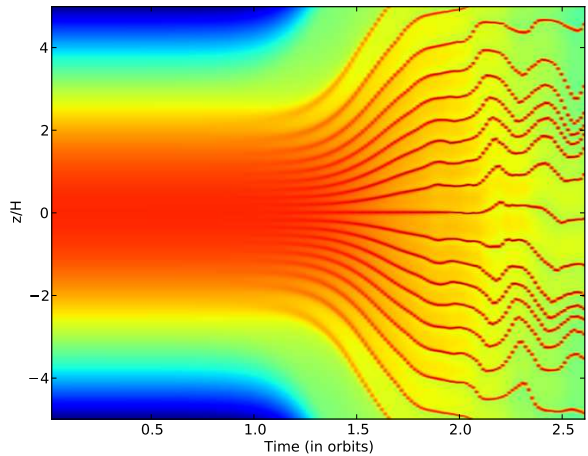}
\caption{As in Fig.~\ref{spacetime_n7} but in the case $n=20$ and
$\beta=1000$.}
\label{spacetime_beta1000_n20}
\end{center}
\end{figure}

 We present the results of
two simulations, the first intialised with a $n=20$ mode and the
second with a $n=30$ mode. The growth rates of both are reported in
Table~\ref{growth_rates_eigen}. Similarly to the $\beta=100$ runs,
there is excellent
agreement between the analytical and the numerical $\sigma$; moreover
 the linear
modes hold their spatial profiles up to large amplitudes. The latter point
 is illustrated in the left panel of
Fig.~\ref{bx_eigen_beta1000}, in which the vertical profile of
$B_x/B_0$ is shown ({\it solid line}) at time $t=1.1$ orbit. The
dashed line shows the same function at time $t=0$, appropriately
rescaled in order to match the amplitude of the eigenmode at
$t=1.1$. The two profiles are in good agreement, even though the mode has
grown by about two orders of magnitude, and at $t=1.1$ is nearly one order of
magnitude greater than the background field $B_0$. 
After $t=1.1$, the midplane $\beta$ dips below 10, and the subsequent
evolution mirrors that of the $\beta=100$ runs.
 This nonlinear behavior is
illustrated in Fig.~\ref{spacetime_beta1000_n20}, a spacetime
diagram for the density in the case $n=20$.

When $\beta=10^4$, the fastest growing eigenmode is $n=68$. Due to its
fine-scale structure, a resolution of $N_z=2048$ is required to describe it.
This corresponds to more than 200 cells per scaleheight (which in 3D would
amount to a formidable computational challenge).
As in the previous examples, the numerical growth rate compares well with the
analytic prediction: the former is $0.7398\pm 0.00018$ (with RAMSES), while the latter is
$0.7424$. In the right panel of Fig.~\ref{bx_eigen_beta1000} the channel's
vertical profile of $B_x/B_0$ is plotted when $t=0$ ({\it dashed line}) and when $t=1.3$ ({\it solid
line}), with the $t=0$ profile rescaled. Because the rapid spatial oscillations are difficult to see, we
truncate the figures at $\pm 2 H$. Note that by $t=1.3$ the channel has
grown by some 25 times the background vertical field, but the
agreement between the two profiles remains excellent.
 Discrepancies emerge
first at intermediate $z$. This is because the (depth-dependent) 
plasma beta associated with a channel is somewhat smaller at
a few scale heights and compressible physics
distort the original profile in these layers before the others.
Thus equipartition will first emerge not at the
midplane but at a few scale heights. But this effect is only noticeable at larger
$n$.

\subsection{Legendre polynomial approximation}

\begin{figure*}
\begin{center}
\includegraphics[scale=0.28]{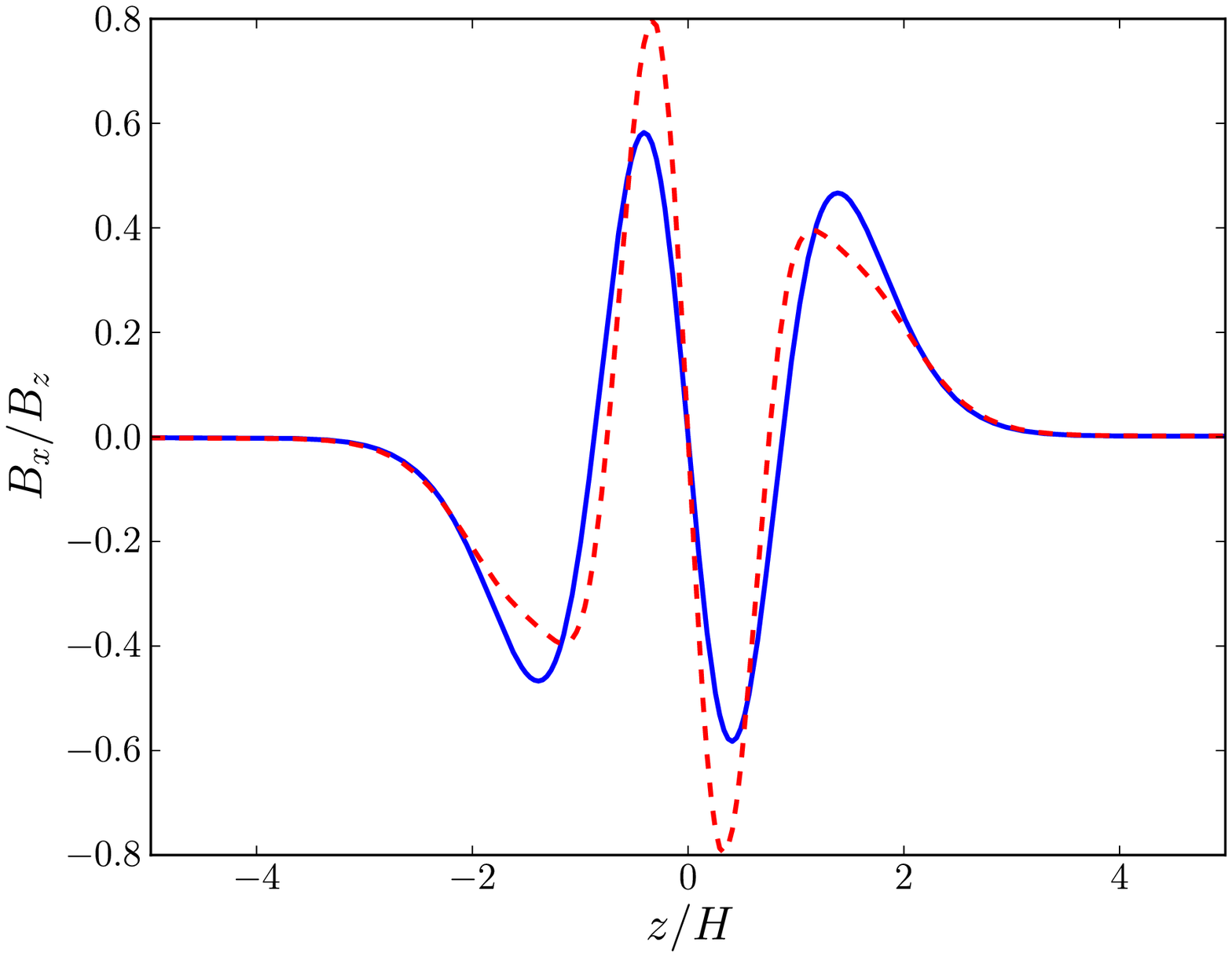}
\includegraphics[scale=0.28]{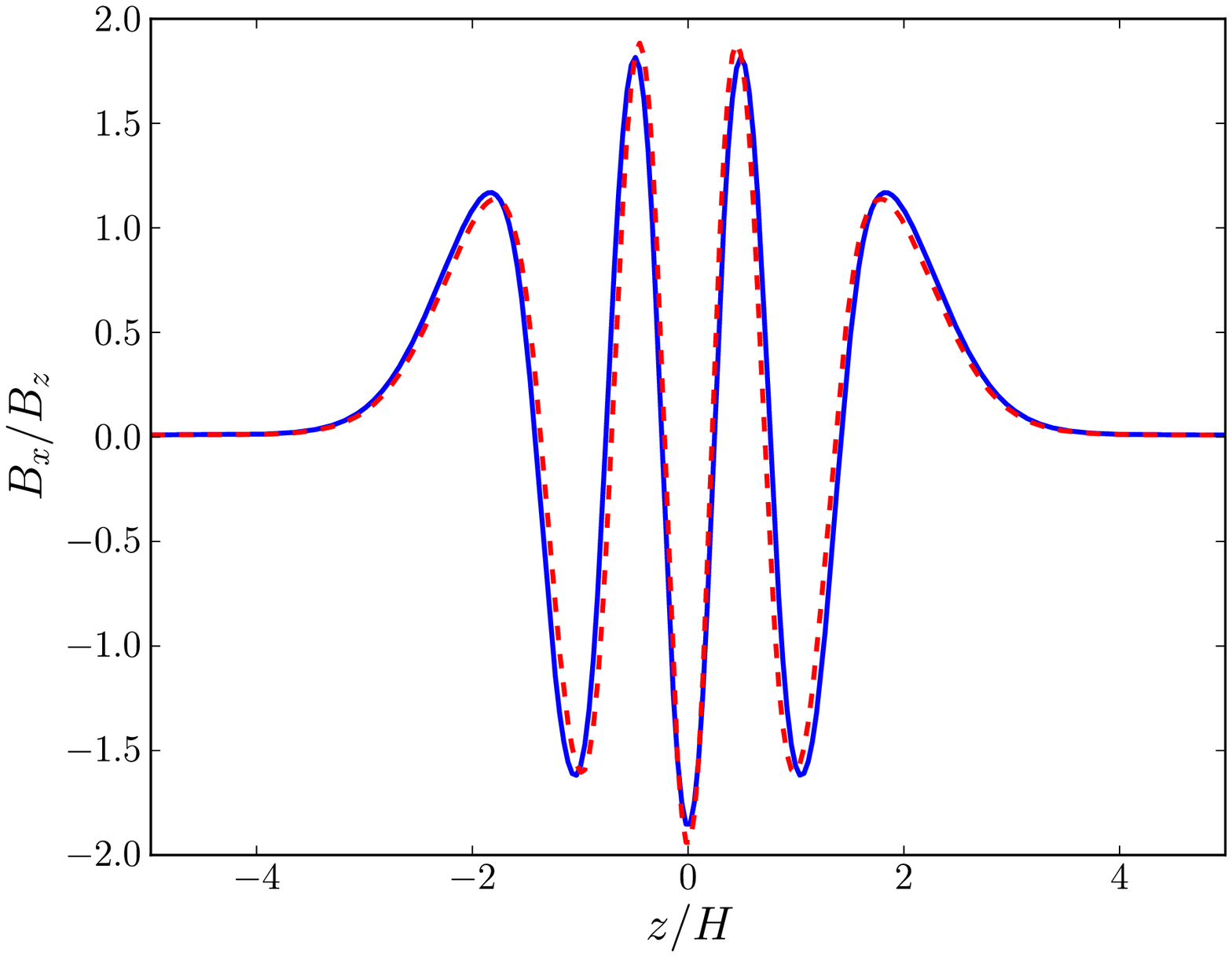}
\includegraphics[scale=0.28]{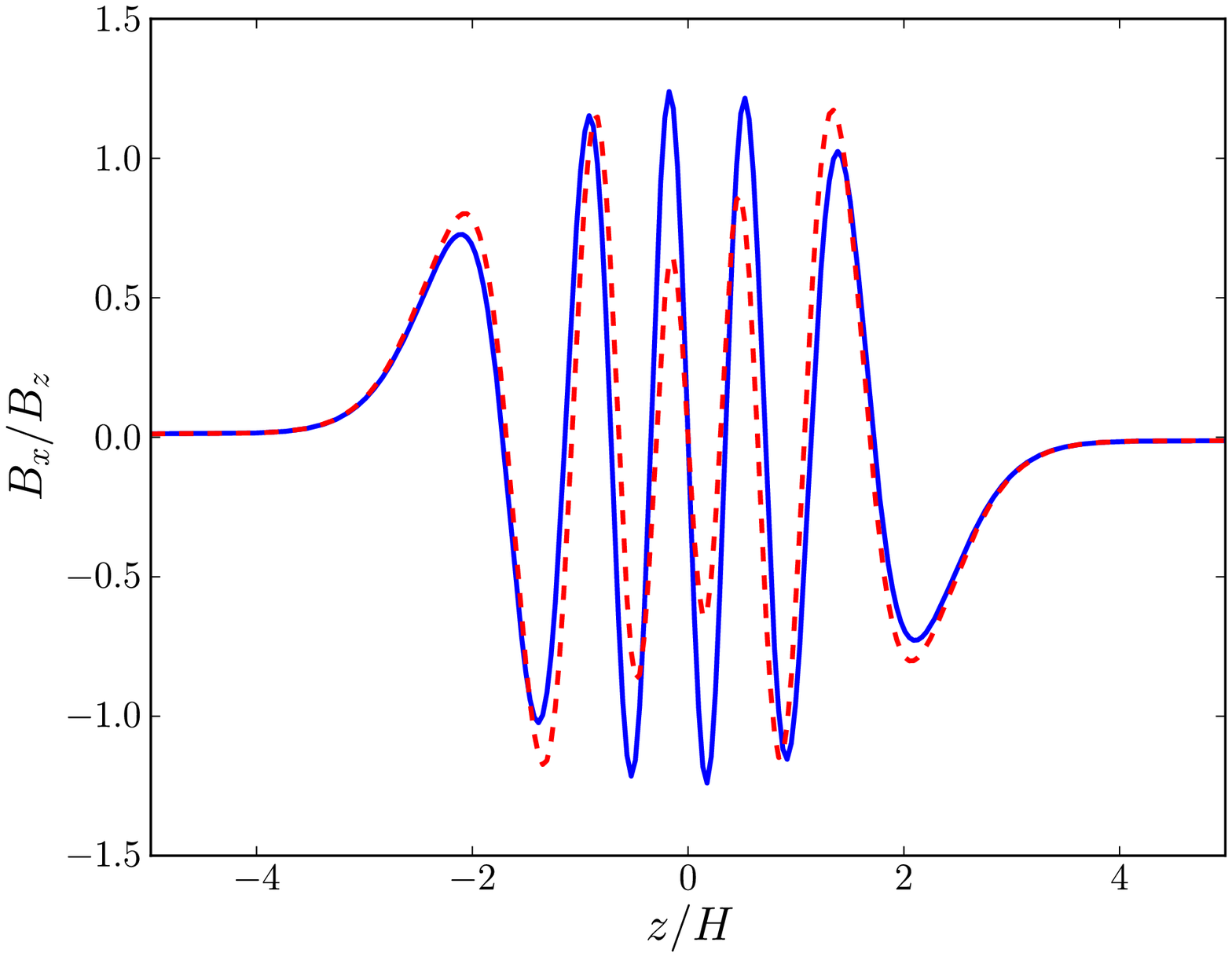}
\caption{In each panel the solid line represents the vertical
profile of the radial component of the magnetic field ({\it solid
line}) after $t=1.2$ orbits for seeded isothermal modes of $n=4$
({\it left panel}), $7$ ({\it middle panel}) and $10$ ({\it right
panel}), all with $\beta=10^2$.
 The dashed lines represent results obtained when the
Legendre approximations to these modes are seeded.}
\label{eignemodes_beta100}
\end{center}
\end{figure*}

We also test the soundness of the approximation introduced in Section 2.4.3.
 To do that, we perform
 simulations in which are seeded the approximate eigenmodes
 of Eq.~\eqref{Legendrefuck},
 though still with the standard Gaussian profile
for the equilibrium density.
For $\beta=10^2$ and $n$ between $4$ and $10$ we find the approximate 
modes hold their vertical profiles relatively well throughout the
 time evolution. This is illustrated in Fig.~\ref{eignemodes_beta100}, which
compares the vertical profile of $B_x/B_0$ obtained with RAMSES at
time $t=1.2$ orbits in simulations in which the exact eigenmodes are
seeded ({\it solid line}) with the results obtained when the Legendre
approximation is seeded ({\it dashed line}). The three panels, from left
to right, respectively correspond to the case $n=4$, $n=7$, and $n=10$. The
 $n=7$ mode is the fastest growing and it offers excellent agreement
between both solutions.
 The agreement is acceptable in
the other two cases, though not so striking. The discrepancy here
 follows simply from the smaller growth rates of these modes. The Legendre
 profiles, being approximations, possess small components in the
 `eigendirections' of faster growing exact modes. Over time, these small
 perturbations, enjoying
  larger growth rates, will distort the evolution of the seeded
approximation, resulting in the imperfect first and third profiles of
Fig.~\ref{eignemodes_beta100}. Nevertheless, these deviations
are controlled and fail to modify the growth rate significantly, as shown
 in the third part of
Table~\ref{growth_rates_eigen} for the modes $n=4$, $7$ and $10$. For
values of $n$ smaller than four and larger than ten, we found that
such problems become important and the Legendre approximation breaks
 down. This is a problem for larger $\beta$, where there are more
 growing modes at larger $n$.
At any given time, however, the set of 
Legendre polynomials offers a useful approximate spectral decomposition of any stratified
 flow into its component MRI modes. This point will be explored in more detail
 in a forthcoming numerical paper.

\section{Parasitic instabilities}

 Stratified channel flows will be subject to
secondary `parasitic' instabilities (GX94, LLB09, Pessah and
 Goodman 2009, Pessah 2009), which can feed on
 their strong velocity shear, and if resistivity is present,
 on their magnetic shear (both evident in Figs 1 and 2).
This section concerns the behaviour of these secondary
instabilities, characterising them mathematically
 in a convenient intermediate limit of channel amplitude. 
 Their role in the destruction of MRI channels is then discussed, as well as
 their importance in sustained MRI-induced turbulence.

\subsection{Intermediate asymptotic limit}

The parasitic analysis is complicated by the fact that the equilibrium
state that we are perturbing varies with $x$ and $t$ (in addition to $z$).
In shearing coordinates the $x$ dependence can be removed, but the time
dependence remains: the channel flow
grows exponentially like $\ee^{st}$, and the background linear shear flow, 
issuing from Keplerian rotation, will `shear out' non-axisymmetric
disturbances over time. 

To get around these
difficulties, we assume that the parasitic modes grow on a timescale much
faster than the channel growth time and the Keplerian shear time (GX94). This,
in fact,
corresponds to an assumption concerning the amplitude of the channels
themselves, and is equivalent to $b\gg 1$.
So the channels must take large amplitudes --- much larger than the
amplitude of the vertical equilbrium field $\B_0$. But if we want to use the
convenient channel profiles computed in Section 2, these amplitudes cannot
be too big, for if they were, then the ansatz, Eqs
\eqref{F}-\eqref{P}, would fail to hold. More
precisely, the midplane Alfven speed associated with the channel $v_A^\ch$
 must be much bigger than the midplane Alfven speed of the vertical background
 $v_A$, and concurrently much smaller than the sound speed $c$. This leads to
 the restriction 
\begin{equation} \label{interm}
 1 \ll b \ll \sqrt{\beta}.
\end{equation}
Obviously, in order for this intermediate range of amplitudes to exist, the initial $\beta$ must
be sufficiently large.
Once this assumption is made, we can use the vertical channel 
structures given by Eq.~\eqref{master},
 while neglecting their exponential growth, the
 Keplerian shear, Coriolis force, and vertical background field. These simplifications
 transform the channel computation into a 1D eigenvalue problem
 in $z$. 

\subsection{Governing equations}

We perturb the channel solution with a small incompressible disturbance, so that
$$ \u= \u_0 + \u^\ch + \u', \quad \B=\B_0 + \B^\ch + \B', \quad \Psi= \Psi_0+ \Psi',$$
where $\Psi$ denotes total pressure and the prime designates
the parasites. For simplicity we have neglected the small variations in the
total pressure
introduced by the channel itself. We have also ignored buoyancy effects
associated with the background vertical pressure structure, which
will have a small stabilising influence on the channel's strong velocity shear
(Drazin and Reid 1982). Together with the incompressibility assumption,
these approximations are particularly valid when $K_n$ is large, which
corresponds to the regime of interest. 
The disturbances are substituted into the governing equations and we linearise in
the small parasitic disturbances. We next assume that $b$ is
large and that the time dependence of the parasites scales like $\d_t\sim
b\Omega$. This allows the dropping of the background Keplerian shear $\u_0$,
Coriolis force,
 and the exponential growth of the channel to leading
order. Throughout most of the domain, this also means that the background
vertical field $\B_0$ is also subdominant, except far from the midplane where
$\B^\ch$ goes to zero. However, it can be shown that the contribution of the
background vertical field in these regions makes little
difference to the overall solution.
 Consequently, as
far as the parasitic mode is concerned, the equilibrium is to leading order
\emph{stationary}, and we are permitted to take normal
modes:
$$\u',\,\B',\, \Psi' \propto \ee^{\ii k_x x + \ii k_y y +
  \sigma t}\,, $$
where $k_x$ and $k_y$ are (real) wavenumbers and $\sigma$ is a growth rate.
 Next, we adopt space and time  units so that
  $H=1$, and $b\Omega=1$, while scaling the velocity, magnetic field, and
  pressure by $b u_0$, $b B_0$, and $b^2 u_0^2 \rho_{00}$ respectively. This
  scaling removes the parameter $b$ from the problem directly. Also,
  we solve for linear momentum density $\pp'$, defined through
\begin{equation}
\pp' = h(z)\,\u', 
\end{equation}
in place of the velocity.
The linearised equations for $\pp'$, $\B'$, and $\Psi'$ are now:
\begin{align}
\sigma \pp' &= -M\left[ p_z'\,\d_z \hat{\u} + \ii(\kk\cdot\hat{\u})\,\pp'
  +(\ii\kk+\ez\d_z)\Psi' \right] \notag \\
&\hskip2cm +\frac{2}{\beta M}\left[B_z'\d_z \hat{\B} + \ii(\kk\cdot\hat{\B})\,\B'
  \right], \label{upar}\\
\sigma \B' &= M\left[B_z'\d_z\hat{\u} - p_z'\,\d_z\hat{\B}/h +
  \ii\pp'(\kk\cdot\hat{\B})/h -\ii\B'\,(\kk\cdot\hat{\u}) \right], \label{Bpar} \\
0 &= \ii \kk\cdot\pp' + \d_z p_z' - \d_z(\ln h)\,p_z', \label{contpar}
\end{align}
where $\kk=k_x\ex+k_y\ey$, the channel background is represented by the `hatted' vectors:
\begin{align*}
\hat{\u}= F_n(z)\,\left(\ex\,\cos\theta + \ey\,\sin\theta \right), \\
\hat{\B}= G_n(z)\,\left(\ex\,\sin\theta -\ey\,\cos\theta \right),
\end{align*}
and where we have introduced the Mach number 
\begin{equation} \label{Mach}
M = u_0/(H\Omega)
\end{equation}
which can be computed from Eq.~\eqref{u0}.
 Equations \eqref{upar}--\eqref{contpar} describe a
one-dimensional eigenvalue problem with eigenvalue $\sigma$. The boundary
conditions require that all variables go to zero at the disk's vertical
boundaries. 

The inputs for the eigenproblem include: the function $h$ (representing the
equilibrium density stratification), the parameter $\beta$ (representing the
strength of the vertical magnetic field), the integer $n$ (which fully
defines the channel flow),
 and $k_x$ and $k_y$
(which define the horizontal variation of the parasitic mode). In place of the
last two it is convenient to use $k=\sqrt{k_x^2+k_y^2}$ and $\theta_k=
\text{arctan}(k_y/k_x)$. 
 These are just the horizontal wavevector magnitude
and its orientation with respect to the $x$ and $y$ axis.
It is also helpful in some circumstances to use
$\Delta=\theta-\theta_k$, which measures the difference in orientation between
a parasite's wavevector and the
direction of the channel flow.
 All the other
quantities and functions that appear in the equations --- $K_n$, $F_n$, $G_n$, $\theta$, $M$ --- depend
solely on $h$, $\beta$, and $n$, through Eqs \eqref{master}, \eqref{theta}, and
\eqref{u0}.

\subsection{Asymptotic analysis}
The full set of equations are
 solved numerically for general parameters, but
some analytic expressions can be obtained in the limit of small $k$. These provide a check on the
numerics and also offer some physical insight. The details of the
analysis can be found in Appendix A; in this subsection we briefly summarise
some of the results pertaining to the fastest growing parasitic mode.

As in GX94 and LLB09, the parasites can be classed into two main types. There is the Type 1,
 Kelvin-Helmholtz or kink, mode which attacks every jet in the sequence of MRI channels,
vertically buckling each in the familiar
sinusoidal pattern. Then there are the Type 2, or kink-pinch, modes, which
 kink some of the jets, but
attack others via an alternate vertical squeezing and
stretching, in a pattern something akin to a sausage (see LLB09). The latter `hybrid'
 modes always grow slower than the kink mode, usually by an order of
 magnitude. Because such modes will rarely destabilise a channel, and their
asymptotic analysis is quite involved, we do not discuss them
 in this subsection. 

In the limit of very long horizontal wavenumber $0<k\ll 1$, the leading order
features of the kink mode are relatively easy to ascertain. Its growth rate is
given by
\begin{equation} \label{kinksigm}
\sigma^2 = A_n\,k^2\,M^2 \left[ \cos^2\Delta- M_A^{-2}
  \sin^2\Delta \right],
\end{equation}
where $A_n$ is a positive number dependent on the channel profile (see Appendix
 A), and
 the Alfvenic Mach number is
$M_A= u_0/v_A^0= M\sqrt{\beta/2}$. The latter depends on $\theta$, but it will
 always be less than 1.
The growth rate scales like $k$, and the modes will stabilise when
\begin{equation} \label{stabcon}
|\theta-\theta_k| > \text{arctan} M_A.
\end{equation}
The stability criterion above is exactly that of parasitic kink modes in the
unstratified case (GX94, LLB09). The kink mode favours orientations
of its wavevector near the direction of the channel flow; the more $\kk$
points away from $\u^\ch$, the less shear energy is available and the greater
the stabilising magnetic tension. When the angle difference
$|\theta-\theta_k|$ is larger than some number less than $\pi/4$ the mode
stabilises.
Conversely, the fastest growing modes are oriented
parallel to the channel, $\theta=\theta_k$, and being perpendicular to the
magnetic field, are essentially hydrodynamic.

\subsection{Numerical solutions}

The Eqs
\eqref{upar}--\eqref{contpar} are solved numerically via a pseudospectral
method similar to that employed in LLB09 but using Whittaker cardinal
functions (as in Section 2.4.2) rather than Fourier modes. 
The Whittaker basis is convenient because it ensures all solutions obey the decaying 
boundary conditions far from the
midplane (Boyd 2001). If $N$ functions and a (uniform) grid spacing of $\delta
z$ are employed,
the governing equations reduce to a
 generalised algebraic eigenvalue problem of order $7(N/\delta z)$, the
 spectrum (or some portion of the spectrum) of
 which may be obtained by the QZ algorithm, or an Arnoldi
 method (Golub and van Loan 1996). Setting $N=16$ and $\delta z=0.04$ sufficed
 in most cases, though these must be improved upon if one is to
 resolve solutions when $n$ is large (because of the channels'
 and their parasites'
 finer spatial structure).  In order to simplify the problem we assume
 that the disk and the MRI modes can be represented adequately by 
 the approximate isothermal model of Section 2.4.3.

As in LLB09, the eigenproblem is subject to many parameters and exhibits a
 rich variety of mode behaviour. In particular, when $n$ increases the number
 and complexity of the set of unstable modes is striking. However, we may
 always decompose this suite of instabilities into (a) a single `pure' kink
 mode, and (b) a subset of slower growing
 (and more physically complicated) kink-pinch modes. To ease their
 presentation, we restrict ourselves to hydrodynamical modes only; that is,
 we set $\theta=\theta_k$. In practice these turn out to be the fastest
 growing, and therefore the most dynamically significant.
 For the same reason, we concentrate our attention primarily on the
 kink mode, and do not catalogue the physical intricacies of the many
 kink-pinch modes that emerge: they just grow too slowly. Finally, we
 specialise to the case $\beta=1000$. This value permits only a marginal
 satisfaction of Eq.~\eqref{interm}, but facilitates the
 calculation.
 In any case, the general structure of the solutions
 are relatively impervious to $\beta$. This leaves the two parameters $n$ and
 $k$.
 When $\beta=1000$ there exist growing
 channel flows for all $n\leq 38$, with the fastest growing channel possessing
 $n=21$. We present detailed results for only two cases $n=2$ and $n=10$, as these
 summarise neatly the overall behaviour. In both cases, $k$ is allowed to vary
 smoothly from 0 to near $K$, at which point the parasitic modes
 stabilise. The results for the $n=21$ mode are complicated, but are
 similar overall to the $n=10$ case.

\begin{figure}
\scalebox{0.55}{\includegraphics{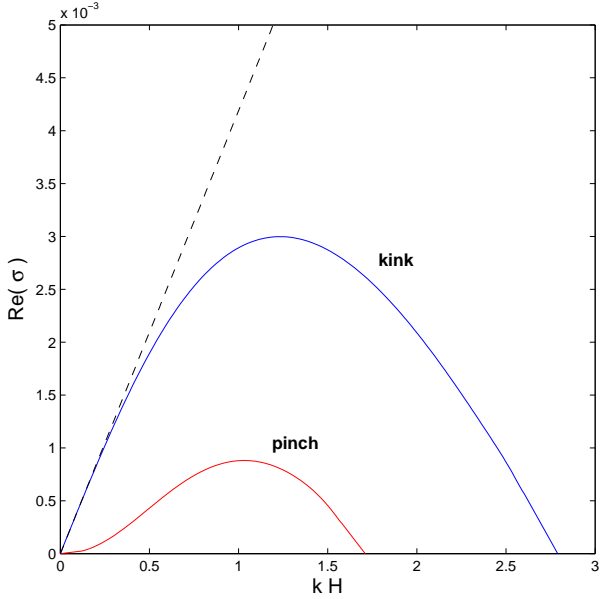}}
\caption{
Growth rates of the unstable kink and pinch mode attacking the $n=2$
MRI channel, which is a single magnetised jet centred at $z=0$. The plasma
beta $\beta$ is set to 1000. The growth rates are
plotted as functions of $kH$, while $\theta_k =\theta$. The latter restriction
means the modes are
essentially hydrodynamic. The asymptotic estimate of $\sigma$ for the kink
mode, cf. Eq.~\eqref{kinksigm}, is plotted with
a dashed line.}
\end{figure}

\subsubsection{The  $n=2$ MRI channel flow: a single jet}

\begin{figure}
\scalebox{0.55}{\includegraphics{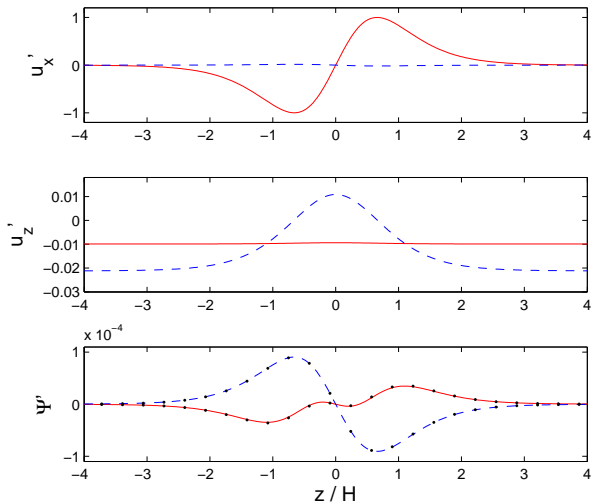}}
\caption{The $u_x'$, $u_z'$ and $\Psi'$ eigenfunctions of a kink mode attacking the
  $n=2$ channel. Here $k=0.01 K (= 0.0245)$, $\theta_k=\theta$, and $\beta=1000$.
 The solid lines indicate
  real part and dashed lines indicate imaginary part. 
 For comparison, the black points represent
the asymptotic eigefunction, cf. \eqref{n2assp}.
 The growth rate of the mode is $\sigma/(b\Omega) =1.03\times 10^{-4} $.}
\end{figure}

The somewhat special $n=1$ channel flow is a simple shear layer (see Fig.~1),
and, as expected, yields only one parasitic mode. This is the classical
Kelvin-Helmholtz instability, which achieves its fastest growth when $k\approx
K/2$ and is extinguished when $k\gtrsim K$ (for $n=1$, $K=$1.1584).
 The $n=2$ channel flow consists of a
single jet centred on the midplane, and is in fact a close relative
 of the $\text{sech}^2 z$ `Bickley jet' (Bickley 1937). Consequently, it is
 subject to two parasitic instabilities, the classical kink and pinch (see
 Drazin and Reid 1981, Biskamp et al.~1998, LLB09). The growth rates of each as
 a function of $k$ are plotted in Fig.~7. As is typical, the growth rate of
 the kink mode is significantly larger than the pinch mode,
 and extends over a wider range of $k$, up to roughly $K$ itself (where, for
 $n=2$, $K=2.0796$). For small
 $k$, the kink mode growth rate $\sigma$ scales as $k$, while the pinch
 appears to go
 as $\sigma\sim k^3$. We also plot in Fig.~7
 the asymptotic prediction
 \eqref{kinksigm} with a dashed line. At low $k$ it offers excellent agreement
 with the computed numerical values.

\begin{figure}
\scalebox{0.55}{\includegraphics{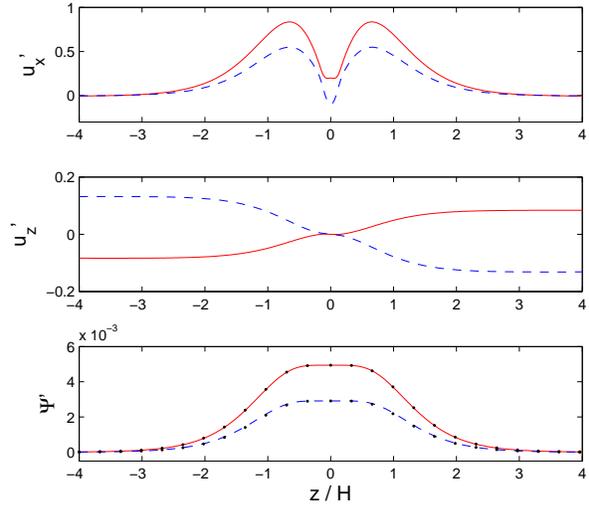}}
\caption{The $u_x'$, $u_z'$ and $\Psi'$ eigenfunctions of a pinch mode upon the
  $n=2$ channel. Here
  $k=0.05 K\, (= 0.12247)$,\, $\theta_k=\theta$, and $\beta=1000$. The growth rate is 
 $\sigma/(b\Omega)=2.49\times 10^{-5}+i\,5.575\times 10^{-4}$, the imaginary
  part of which is in reasonable agreement with the leading order asymptotic
  theory, which gives $\sigma= i 5.75\times 10^{-4}$. The black points
  represent the leading order $\Psi'$ asymptotic profile.}
\end{figure}

Typical profiles of the two modes are presented in Figs 8 and 9 at small $k$,
alongside the asymptotic pressure profiles. These low $k$ modes should rarely be
observed because they grow too slowly, but they allow a clearer 
demonstration of the main physics and a clean comparison with the asymptotic theory.
 The kink mode in the first
figure is easy to distinguish because it possesses a vertical velocity maximum
at the centre of the jet, $z=0$. Correspondingly, the pressure pertubation at
this point exhibits a strong vertical gradient. As a consequence, the mode elicits an
alternating horizontal sequence of upward and downward motions along $\kk$, which will
buckle the jet in the familiar Kelvin-Helmholtz pattern.
 In contrast, the pinch mode in Fig.~9
exhibits a pressure maximum at the jet centre, while $u_z'$ changes sign. This
corresponds to a horizontal sequence of squeezes and rarefactions as fluid is
either drawn towards or repelled by the jet.

Other $\theta_k$ were tried, but only a very narrow range of orientations
(around $\theta$)
permitted growing modes. As predicted by the criterion \eqref{stabcon}, the
kink mode stabilises when $|\theta_k-\theta|$ exceeds about 12 degrees. The
interval is particularly small because the $n=2$ channel is very
sub-Alfv\'enic for $\beta=1000$. Faster growing (larger $n$) channels yield greater $M_A$
and, consequently, a greater range of unstable $\theta_k$. In common with its
kink-pinch counterparts in an
unstratified disk, the pinch mode grows for a broader range of $\theta_k$ (GX94,
LLB09, Pessah and Goodman 2009, Pessah 2009).

Finally, note that these instabilities will fail to appear in numerical
simulations with short radial domains, such as one scale height $H$. Because $K_2
\approx 2.45/H$, the radial box length should be at least $2\pi/K_2\approx
2.57\,H$ to capture the shortest horizontal modes. As $n$ increases for given
$\beta$ this constraint becomes less of a problem: for $n=10$ a radial length
of $H$ will comfortably fit a number of growing kink modes.

\subsubsection{The $n=10$ MRI channel flow: 9 jets}

Larger $n$ channels support a greater number of unstable parasitic modes; this
 is simply because larger $n$ channels' exhibit more complicated vertical structures
 and thus permit a wider array of destabilising
configurations. The presentation of these many modes at
 various
 $n$ is too lengthy and laborious a task for this paper. Instead we show results
 from a representative higher order channel, associated with $n=10$. 

From Fig.~1, the $n=10$ channel flow consists of a jet situated at the
midplane $z=0$ and a symmetric set of 4 jets on each side. We find that these
can harbour up to 9  parasitic modes. These include the kink mode, which grows
the fastest and kinks every jet, and 8 kink-pinch modes, which differ in the
number and location of their component kinks and pinches. Not every
combination of kinks and pinches are supported by the system (if this was so
there could be $\sim 2^9$ such modes!). One important constraint is that
no two pinches can attack adjacent jets, a limitation also evident in the
unstratified problem (which admits no `pinch-pinch' modes, LLB09). Some
modes also leave a number of jets unmolested, with $u_z'$ both a maximum and
zero at these locations. When $k$ takes larger values this becomes more
noticeable, and the fastest growing modes attack those jets nearest
the midplane preferentially.

\begin{figure}
\scalebox{0.55}{\includegraphics{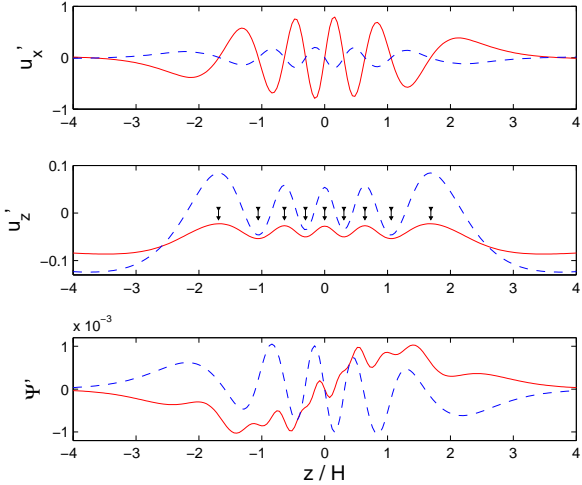}}
\caption{Selected components of the kink mode eigenfunction, for parameters:
  $n=10$, $\beta=1000$, and $k=0.05\,K\,(=0.524)$, $\theta_k=\theta$.
 The growth rate of the mode is $\sigma/(b\Omega)=3.01\times 10^{-3}$. The short
  vertical arrows in the central panel indicate the locations of the $n=10$ MRI
  channels. }
\end{figure}

We plot the eigenfunction of the `pure' kink mode in Fig.~10 for small
$k$. Again, we choose small $k$ modes as they demonstrate the physics the
clearest. We deal with the fastest growing modes in the following subsection.
 In the
central panel we also indicate the centres of the nine channels by short
vertical arrows. These show that the local extrema of the parasite's vertical velocity
are located exactly at the channel centres. Thus the mode kinks or buckles
every jet concurrently. Simultaneously, the channel centres
correspond to those locations where the pressure perturbation
possesses a large gradient, though this is less easy to see. Far clearer,
 is the large-scale antisymmetry in the pressure perturbation,
especially in its real part. This means that in addition to the localised
kinks upon each channel there exists a `global' kink upon the \emph{entire} set of
jets as well. Variations in the eigenfunctions' large-scale `envelope' are a
feature of the more complicated high $n$ parasites. In the unstratified
analysis, the larger scale structure manifests
 simply through the Floquet factor $\text{e}^{\ii k_z z}$ (see GX94 and
LLB09). 
The maximum growth rate for the kink mode (when $n=10$) is $\sigma=0.01726$ achieved when
$k=0.56K$. The mode stabilises when $k\geq 0.97K$. (Note that for $n=10$, we
have $K=9.2239$.)

\begin{figure}
\scalebox{0.55}{\includegraphics{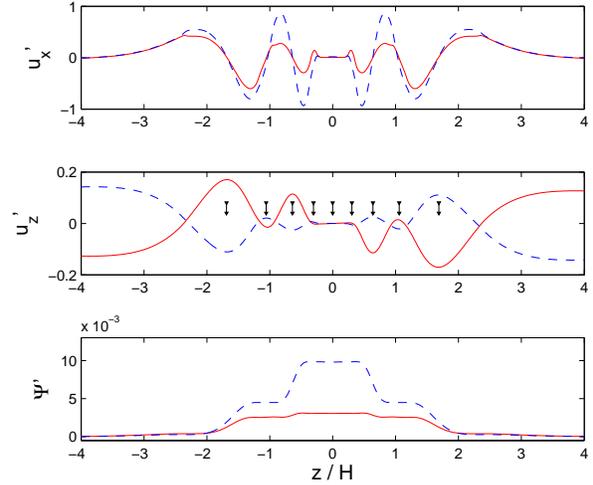}}
\caption{Selected eigenfunction components of a kink-pinch mode, with
  parameters as in Fig.~10. The growth rate is $\sigma/(b\Omega)= 5.84\times 10^{-4}-
  i\,3.18\times 10^{-3}$. As earlier, the black vertical arows indicate the
  locations of the MRI channels. Here the central three channels are pinched
  as a whole and the
  remaining channels are kinked. }
\end{figure}

We present one example of a kink-pinch eigenfunction in Fig.~11. As earlier, the
small vertical arrows indicate the locations of the channel centres. The
vertical velocity possesses local extrema at all the centres except the
jet at the midplane. At the midplane the vertical velocity
changes sign, which means this mode pinches
the central jet. The other 8 jets, in contrast, are kinked (to varying degrees). The pressure pertubation tells us,
on the other hand, that the mode exhibits a large-scale scale envelope through which the
entire set of jets is pinched. This is also clear from the large-scale antisymmetry in the
$u_z'$ profile.

\subsubsection{Fastest growing modes}

 The maximum growth rates of the kink mode
 computed with our stratified model are significantly less
  than growth rates in the unstratified model (GX94). 
 The unstratified parasitic kink mode reaches a maximum 
 $\sigma \approx 0.2\,b\Omega$ when attacking the fastest growing MRI
 channel.
 In contrast,
\begin{itemize}
\item when $\beta=10^2$, the
 $n=7$ MRI channel is dominant and its fastest growing parasite possesses
 $\sigma=0.0596\,b\Omega$, at $k\approx 0.525\,K_7$. 
\item When $\beta=10^3$, the $n=21$ channel is dominant
 and the maximum $\sigma$ is $0.0329\,b\Omega$, at $k\approx 0.575\,K_{21}$.
\item And when $\beta=10^4$, the $n=68$ channel dominates and
$\sigma_\text{max}=0.0192\,b\Omega$ at $k\approx 0.585\,K_{68}$.
\end{itemize} 
 This discrepancy is something of a surprise. One may have expected that at
 sufficiently large $\beta$ the unstratified and stratified results would
 converge. Instead there appears a scaling, with possibly the logarithm of
 $\beta$, which diverges from the unstratified growth rate.
 The difference in $\sigma_\text{max}$ probably results from the 
 constrasting
 global structure of the channel flows. While an unstratified parasite
 attacks an
 infinite lattice of channels of equal amplitude, 
its stratified cousins can only attack a \emph{finite} number of channels, of varying
 amplitude.
Consequently, the amount and configuration of the free energy available are
 dissimilar, and obviously influence the maximum growth rates that
 parasites can achieve.   

Because the fastest growing modes have large $k$, the eigenfunctions tend to
localise near the midplane where the velocity shear is greatest. They also
lose any large-scale envelope they may have exhibited. This can be observed in Fig.~12
which presents the fastest
growing mode on the $n=21$ channel when $\beta=1000$. Here the mode is individually kinking
the middle 12 jets, and ignoring the other 8 at larger altitudes. Because
these central jets are more closely packed they offer greater velocity shear
for the parasitic mode. This behaviour is typical for the fastest growing modes. 
It follows that, if they can be adequately resolved, parasites will be found
primarily near the midplane in numerical simulations.

\begin{figure}
\scalebox{0.55}{\includegraphics{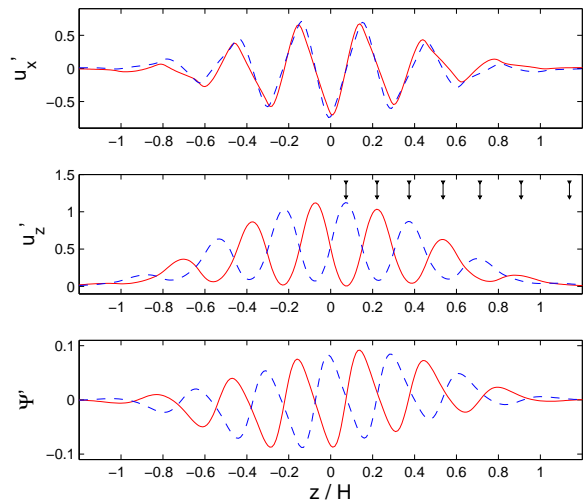}}
\caption{Selected eigenfunction components of the kink mode upon the $n=21$
  channel for $\beta=1000$. The horizontal wavenumber is
  $k=0.575\,K\,\,(=12.36)$ and $\theta_k=\theta$, thus
  the growth rate is $\sigma/(b\Omega)= 0.0329$. The black vertical arrows indicate the
  locations of the MRI channels in the range $1.2>z>0$, a symmetric set occurs for
  $z<0$ but is omitted (there are 20 altogether). Note that the mode has localised upon those
  channels near 
the midplane, while ignoring those at larger altitudes. This is typical of
  larger $k$ modes, and hence the fastest growing parasites.}
\end{figure}

A final point concerns the consistency of the intermediate limit we
employ. The $\sigma_\text{max}$ of the stratified parasites are not only
smaller than their unstratified counterparts, they are sufficiently small to
endanger the original scaling $\sigma\sim b\Omega$. This means one should
revise the lower limit in Eq.~\eqref{interm}, which defines the
intermediate regime, possibly increasing it to 10 or even 100. But unless $\beta$
 is very large indeed ($>10^5$, say), there may be no intermediate regime
 to speak of:
 one must either account for the background shear and channel growth,
 or one must include compressibility. Having said that, the general features
 of the analysis --- the estimates of the growth rates and the modes'
 general structurem --- should offer a satisfactory guide to the parasitic
 physics at any $b$. In the next subsection we discuss the more realistic
 non-intermediate regimes.

\subsection{Discussion}

Our treatment of the stratified parasitic modes provides only a limited survey
of this problem's rich assortment of mode behaviour. But the
salient dynamical points remain clear: the fastest growing mode is always
the hydrodynamical kink mode (with $\theta_k=\theta$), favouring a wavenumber
some half that of its host channel $K$. This is also the case in the
unstratified analysis. While there exist an interesting array
of other modes, exhibiting complicated morphologies, these pinch and kink-pinch modes
 grow significantly slower, and so should be
 dynamically subdominant. In short, if a channel is to be disrupted by a
 parasite, it will be the kink mode doing the disrupting. Finally, the maximum
 growth rates that stratified parasitic modes attain are 
significantly less than those calculated from unstratified models.

\subsubsection{Small amplitude channels}

Our mathematical
characterisation of the parasitic modes only holds in an intermediate limit of channel
amplitude: channels must be sufficiently large that we may omit the Keplerian shear and
the channels' exponential growth, but sufficiently small that
compressible effects are negligible. However, in order to understand the role of
the parasites in the MRI dynamics more generally, we need to
extrapolate these intermediate results into the regimes of small and large
channel amplitudes.

An important dynamical question is: How large will an MRI channel become
before it is destroyed by a parasite? Will a channel reach equipartition
before collapsing? This is important not only to understand (and possibly
control) the MRI's initial eruptive behaviour, but also to assess whether
continuous channel production and destruction are key ingredients in MRI-induced
turbulence (see Pessah and Goodman 2009, Pessah 2009).
In order to answer this question we must
understand the parasites' behaviour when the channel amplitudes are small.

One could, of course, assume that the intermediate amplitude profiles and growth rates (previously
calculated) offer a reasonable approximation in the small amplitude regime, as do Pessah and
Goodman (2009) and Pessah (2009). This means one expects the exponential growth of
the channel and the Keplerian shear to not qualitatively alter the solutions.
By simply equating the channel growth rate and the
kink mode growth rate, we can obtain a lower bound on the minimum $b$ necessary for
channel disruption. In an unstratified ideal MHD channel this value is roughly
4 (Pessah and Goodman 2009); in the stratified model employed in the previous
section with $\beta=1000$, a similar calculation gives $b_\text{min}\approx
15$. We now argue that both lower bounds are far too small.

In LLB09 the influence of the Keplerian shear and channel growth is discussed in
a qualitative way. In that paper the channel growth was 
emphasised as key in delaying the
onset of the parasitic modes. The role of the shear in doing the same was
not considered.
 In fact, the shear should impede the parasites equally efficiently. The Keplerian
shear introduces a time-dependence to the parasitic wavevector $\kk$. As time
progresses the
orientation angle $\theta_k$ will decrease and the wavenumber $k$ will
increase. But parasitic modes only grow when their $\theta_k$
 lie within some range encompassing $\theta$ and when $k<K$ (see
 Eq.~\eqref{stabcon} and Fig.~7, for example). It follows directly that
 these modes will grow only transiently, enjoying a short burst as
 $\theta_k(t)$ and $k(t)$ pass through their respective neighbourhoods of
 permitted growth. If the mode does not grow sufficiently fast within this finite
 time interval --- sufficiently fast to reach an amplitude comparable to $b$ --- then the channel
 remains undisrupted. This time restriction provides a means to calculate the
 minimum $b$ necessary for channel disruption. We present these details in
 Appendix B. There we find that for the unstratified problem, and for an initial
 parasitic amplitude $1/10$ or $1/100$ of the channel, the critical amplitude is $b_\text{min}\approx
 24$ or $40$. In the
 stratified problem, these values are of order $100$. Both estimates are
 considerably larger than if the Keplerian shear is omitted. Both also show
 that channels, particularly stratified channels, are destined to achieve
 large amplitudes and will likely hit equipartition before the parasites can
 topple them: a channel that has grown from $b=1$ to a hundred times that,
 will have decreased the midplane $\beta$ by a factor $10^4$.

 \subsubsection{Compressible channels}

Once the system has reached equipartition the analyses of Sections 2 and
4 break down. The profile of the MRI channel will alter, its magnetic
pressure sufficiently strong to squeeze mass into its jet centres 
(see Figs 3 and 5 and Stone and Miller 2000). 
The channel jets will become thinner and more intense, while 
simultaneously coinciding with narrow current sheets. If sufficiently thin, the
influence of resistivity will become important, giving rise to new parasitic
tearing modes (explored in LLB09). Finite resolution will ensure that the magnetic
reconnection induced by these modes features prominently in stratified box simulations.
Compressibility may introduce parasitic Parker modes, which will exploit the
strong bands of magnetic field between jet centres. The character
of the original kink modes will change as well, though perhaps not
drastically. As shown in LLB09, when the jets narrow and become more intense,
 the kink
mode tends to grow on shorter horizontal length scales, and can split
 into multiple modes each localised to a single jet. This new wealth of
 dynamical behaviour should be observable in resolved 3D stratified simulations, and we
 intend to present some of this work in a later paper.

\section{Conclusion}

In summary, we have shown that channel flows, a subset of the linear axisymmetric
MRI modes, are approximate nonlinear solutions in the vertically-stratified
shearing box, when in the limit of large midplane $\beta$. This 
generalises the better known unstratified result (GX94), but only holds when
the equilibrium consists of a net vertical magnetic flux. We compute stratified
channel flows numerically for an isothermal model, and confirm the basic
result with 1D 
simulations. The MRI eigenfunctions and growth rates, conversely, offer a valuable
numerical benchmark on the performance of codes operating in the
vertically-stratified shearing box. In addition, these simulations provide 
estimates (for future 3D simulations) of the 
resolution required to resolve the fastest growing channels when 
using second order numerical schemes: 25 cells per 
scaleheight are needed when $\beta=100$, 50 cells when $\beta=10^3$, and
 more than 200 cells per scale height are 
mandatory when $\beta=10^4$.

Stratified channel flows, like their unstratified counterparts, are subject to
a variety of magnetised shear flow instabilities, or `parasitic modes'. Some
of these we directly compute in an intermediate limit of channel
amplitude. As in the unstratified setting, the fastest growing parasite is the
`unmagnetised' kink mode. But it is unlikely to destroy its host before
 equipartition fields have been reached. 

At present we are extending these findings with 3D numerical simulations in
stratified boxes, where the parasitic modes' behaviour can be confirmed, not
only in the initial stages of a simulation but once it has relaxed into a
saturated turbulent state. Because of the inevitably of large magnetic fields,
and also finite resolution, these simulations will permit a set of interesting physical
 effects which we have neglected here but which deserve further study. These
 include, most importantly, magnetic buoyancy
and magnetic reconnection. Also, these simulations will allow us to test the
recurrent formation of channels once the MRI has saturated in turbulence
(Suzuki and Inutsuka 2009). In
particular, they can establish whether
  larger radial domains impede
recurrent channel formation, by analogy with unstratified boxes (Bodo et
al.~2008, LLB09).

Lastly, we speculate on the potential role dominant channel flows play in real
magnetised accretion disks. Global disk models support
 analogues of the stratified MRI channels, localised in radius
 to narrow rings. Such modes might also exhibit a
 weaker version of the nonlinear property examined in this paper,
 permitting one mode to dominate the initial
 evolution of small amplitude initial conditions. Indeed, 
 simulations with a poloidal field piercing the disk
 report such behaviour (Hawley 2000, 2001). In fully-developed turbulent
 disks, on the other hand, the
 dominance of a single channel will be rare. But occasional `run-away' channel flows
 might occur in the marginally stable fluid that divides regions of stable and
 unstable MRI --- dead-zone interfaces in protostellar disks, for instance. It may
 take only a small fluctuation in the ionisation and thermal properties of the
 gas to ignite instability in this previously quiescent critical layer, and hence launch
 a powerful MRI channel flow. Disruptive events of this kind may control the
 integral-scale intermittency that should characterise the dead-zone
 interface, and may even be related to the initial phases of
 outburst behaviour (Zhu et al.~2009). Similarly, channel
 flows may also feature in
the onset of episodic accretion in dwarf nova disks (Gammie and Menou 1998).

\section*{Acknowledgments}

The authors acknowledge the careful and detailed report by the anonymous
reviewer, which led to a much improved manuscript.
HNL and SF would like thank Alexandros Alexakis, Steven Balbus,
 Tobias Heinemann, Pierre Lesaffre, Geoffroy Lesur,
and Pierre-Yves Longaretti for helpful
advice. HNL acknowledges
 funding from the Conseil R\'egional de l'Ile de France.
SF acknowledges the Isaac Newton Institute for Mathematical Sciences 
for hospitality during August 2009 during which most of the numerical 
simulations presented in the paper were completed. Such simulations were 
granted access to the HPC resouces of CINES under the allocation 
x2009042231 made by GENCI (Grand Equipement National de Calcul Intensif). 
OG would like to thank the CEA Service Astrophysique for its hospitality,
and where some of the numerical simulations were undertaken.

\appendix

\section{Asymptotic analysis of parasitic instabilities}

In this appendix we sketch out an asymptotic solution to
 Eqs \eqref{upar}--\eqref{contpar} in the limit of small horizontal parasitic
 wavenumber: $0<k\ll 1$. We concentrate on a description
 of the kink mode as it is the fastest growing, and hence most dynamically
 significant.

  First we manipulate the governing equations into
single second order equation for the pressure pertubation $\Psi$:
\begin{equation} \label{Lag}
g(z)\,\frac{d}{dz}\left(\frac{1}{g(z)}\frac{d\Psi}{dz}\right) - k^2 \Psi = 0,
\end{equation}
where
\begin{equation}
g(z)= h(z)\left[\sigma+ikM F_n(z)\,\cos\Delta \right]^2 +
\frac{2}{\beta} k^2\,G_n(z)^2\,\sin^2\Delta .
\end{equation}
recall $\Delta=\theta-\theta_k$, which is just the difference in orientation
angles of the MRI channel and its parasite, and $M$ is the Mach number.
The algebra required to derive this can be greatly reduced by using Lagrangian
displacements, not momenta (see GX94). The reader will notice that Eq~\eqref{Lag} is similar
to Eq.~(25) in GX94. We can reproduce this equation if we solve instead for the
Lagrangian vertical displacement and assume the disk is
unstratified. One can derive an analogous `semicircle theorem' for \eqref{Lag}
which shows that both the growth rate and frequency of the parasite is bounded
above by $k$. Thus as $k$ goes to 0, so must both the real and imaginary part of $\sigma$.

\subsection{Kink modes}

We now suppose that $k$ is sufficiently small that we may drop the last term
in Eq.~\eqref{Lag}. This assumption stipulates that the horizontal variation
of the parasite is much greater than its vertical variation, as well as variations in
the channel and the equilibrium density (which are of order $H$ at most).

We can solve directly, in this case, for $\Psi$:
$$ \Psi = c_1 + c_2 \int_{-z_B}^z g(\hat{z})\,d\hat{z} $$
where $c_1$ and $c_2$ are integration constants and $z_B$ is the upper
boundary of the disk (and may be $\infty$). We require that $\Psi=0$ at
$z=\pm z_B$ (the vanishing boundary conditions). This furnishes $c_1=0$ and the eigenvalue equation
$$ \int_{-z_B}^{z_B} g\,dz = 0.$$
After some manipulation we obtain the growth rate explicitly:
\begin{equation} \label{kinksig}
\sigma^2 = A_n\,k^2\,M^2 \left[ \cos^2\Delta- M_A^{-2}
  \sin^2\Delta \right],
\end{equation}
where 
\begin{equation}\label{An}
 A_n= \left(\int_{-z_B}^{z_B} h F_n^2 \,dz\right)/\left(\int_{-z_B}^{z_B} h \,dz
\right).
\end{equation}
 The Alfvenic Mach number
$M_A= u_0/v_A^0$ has been introduced in the above, which can be re-expressed in terms of $M$ and $\beta$.
In deriving equation \eqref{kinksig} the
following two identities were required
$$ \int_{-z_B}^{z_B} hF^2 dz = \int_{-z_B}^{z_B} G^2 dz, \qquad
\int_{-z_B}^{z_B} h F dz = 0,$$
the first follows from multiplying Eq.~\eqref{master} by $F$ and integrating,
the second from the relationship between $F$ and $dG/dz$. 

As anticipated, the growth rate scales like $k$ and will be real (and positive) unless
\begin{equation}
|\theta-\theta_k| > \text{arctan} M_A,
\end{equation} 
which is precisely the stability condition we find in the unstratified case for kink
 modes (see GX94, and LLB09).  The
 criterion just says that if a parasite is to grow, its wavevector must point
 sufficiently along the shear flow.

The eigenfunctions can be obtained analytically if we use the `approximate
isothermal model' (cf. Section 2.4.3). Then $A_n= 1/(2n+1)$, and in the hydrodynamical case of
$\theta_k=\theta$ we obtain
\begin{align}
& \Psi_{1k} \propto \text{sech}^2 z\left(\text{tanh}z-i\sqrt{3}
 \right), \\
& \Psi_{2k} \propto \text{tanh}z\,\text{sech}^4 z \left[ (5-4\,\text{cosh}2z)+
 2i\sqrt{5}(1+\text{cosh}2z) \right] \label{n2assp}
\end{align}
for $n=1$ and $n=2$ respectively. The $\Psi_{2k}$ eigenfunction is plotted in
Fig.~8, and provides a check on the full numerical solution.

Finally note that if the disk is unstratified we can compute $A_n=1/2$ if the
limit of $z_B\to \infty$ is taken carefully in both the denominator and
numerator of \eqref{An} (see also GX). Interestingly, this prefactor is
independent of the mode number or $K$, in contrast to the stratified case
of the previous paragraph where larger $n$ furnishes slower growing modes at
small $k$. According to Section 4.5.1 this seems also to be the case for general $k$.

\subsection{Pinch and kink-pinch modes}

The mathematical derivation of the pinch and kink-pinch modes is a great deal
more involved and subtle, not least because the real part of the growth rate $\sigma$
comes in at higher orders. We will not produce the full analysis
here. Instead, we work out the problem for the simple $n=2$ case to leading order.

The $n=2$ MRI mode is a single jet centred at $z=0$. It follows that there are
two types of parasite: odd (kink modes) for which $\Psi(0)=0$, and even (pinch
modes) for which $d\Psi/dz(0)=0$. At leading order the latter boundary
condition
 gives the equation:
$$ g(0)=0,$$
which is satisfied only if
\begin{equation}
\sigma= -ikM\,F(0)\,\cos\Delta.
\end{equation}
 Applying the decaying boundary condition at $z=z_B$
completes the description of
the pinch mode, the vertical profile of which is
\begin{equation}
\Psi\propto 1- \left(\int_0^z g(\hat{z})\,d\hat{z}\right)/\left( \int_0^{z_B} g\,dz\right)\,, 
\end{equation}
on the domain $z\in [\,0,\,z_B\,]$. If we apply the approximate isothermal model and
furthermore take $\theta=\theta_k$,
then this reduces to
\begin{equation}
\Psi_{2p} \propto 1- \text{tanh}^5 z.
\end{equation}
The eigenfunction is plotted in
Fig.~9 against the numerical solution.
The real part of the growth rate comes in at higher order and requires an additional
asymptotic matching procedure, the details of which we
 do not provide here.

\section{Small amplitude channels}

The purpose of this appendix is to derive an estimate for the smallest
amplitude $b$ that permits parasitic modes to overcome the primary
channel flow. Channels with amplitudes below this critical value host parasites
that grow too slowly to cause disruption.
 We define `channel disruption' when
 the \emph{amplitudes} of the channel and the parasite are equal, in contrast
to when their \emph{growth rates} are equal (Pessah and Goodman 2009, Pessah
2009). In the following, the influence of
the background shear will be included, and we will find that it is
a primary agent delaying channel disruption. We do not treat the
 influence of the Coriolis force.

For simplicity, the disk is assumed to be \emph{unstratified}. Consider the fastest
growing MRI channel, with $s=(3/4)\Omega$ and $\theta=\pi/4$. Its fastest
growing parasite, the one in which we will be most interested, is the kink
mode. It achieves its maximum growth when $\theta_k=\theta=\pi/4$ and $k\approx
0.6\,K$, at which
point $\sigma\approx 0.2\,b\Omega$. It, however, can grow at a reduced rate
within the sector $|\theta-\theta_k|< \theta_\text{crit}<\pi/4$, and for $k<K$
(see GX94). 

Suppose that we allow for the background shear's effect on the parasitic
wavevector $\kk$. Under its influence the wavevector will become
time-dependent, lengthening with time and rotating so that it points
more and more radially.
These changes are summarised by 
\begin{equation} \label{kxt}
k_x(t)= k_x^0 + \frac{3}{2}\Omega\,k_y^0\,t,
\end{equation}
where $(k_x^0,\,k_y^0)$ is the initial value of the wavevector. Note that the azimuthal
component $k_y^0$ is independent of time.

We analyse a fiducial kink mode with $k_y^0= (0.6/\sqrt{2})K$ and
$k_x^0=0$.\footnote[1]{We assume that a kink mode poses the greatest threat to
  channel stability. Though pinch and kink-pinch modes grow for a larger range
of $\theta_k$, they grow for a smaller range of $k$. This combined with their
smaller growth rates leads us to discount them.} Thus the kink mode starts
out with a purely azimuthal wavevector and zero growth rate. It, however, will achieve the maximum
growth possible at some later time when $k_x=k_y$, i.e.\ when $\kk$ is perfectly
aligned with the channel. After passing this `sweet spot' its growth rate will
decline, and when $k_x^\text{f}= \sqrt{0.82}\,K$ it will be zero, because at that point
$k=K$. If we permit the mode to grow between orientations $k_x=0$ and
$k_x=k_x^\text{f}$, we can then estimate the total time the mode grows from
Eq.~\eqref{kxt}. We have, in fact, 
\begin{equation}
t= \frac{2(k_x^f-k_x^0)}{3\Omega k_y^0} \approx 1.42\,\Omega^{-1}. 
\end{equation}
The growth time alotted to
a kink mode is thus roughly a quarter of an orbit. This is the maximum time
available for it to disrupt its host. If it cannot reach a sufficiently large
amplitude within this time then the channel will keep growing unimpeded 
(until another kink mode emerges and tries its luck). Of course, the success of the parasite's attack depends
on the average size of $\sigma$ over the period of growth. We can, in
fact, easily derive a minimum value for this average growth rate, below which a kink
mode cannot disrupt the channel. And because $\sigma$ is proportional to $b$,
this in turn allows us to calculate the minimum $b$ below which channels are
safe from kink modes.

To ease the
calculation we shall take this average value to be half the maximum growth,
i.e.\ $\sigma_\text{av} = 0.1\,b\Omega$. We also must attribute a starting
amplitude to the channel, which we take to be $Rb$, where $R$ is some
fraction. During the time $t= 1.42/\Omega$, the channel will grow by
approximately $\text{e}^{1.07}$, while the kink mode will grow by approximately
$\text{e}^{0.142b}$. The ratio of their amplitudes at the end of this time is
hence
\begin{equation}
\frac{\text{parasite amplitude}}{\text{channel amplitude}}\approx
R\,\text{exp}(0.142\,b- 1.07).
\end{equation}
If this ratio is set to $1$, then we can calculate the critical (minimum) $b$. We have simply
\begin{equation} \label{form}
b_\text{cr} \approx 7.54 - 7.04\,\text{ln}\,R.
\end{equation}
If the initial parasitic amplitude is set to one tenth the channel, i.e.\
$R=1/10$ then the minimum $b$ for channel disruption is about $24$, while
$R=1/100$ gives $b_\text{cr}$ closer to $40$. In either case, channels will grow to
significantly larger amplitudes than predicted by Pessah and Goodman (2009)
and Pessah (2009),
who neglect the influence of the background shear in retarding the parasitic
growth. 

The analysis above was carried out for the unstratified problem, but it is
easy to generalise to the stratified disk. As mentioned in Section 4, the
stratified parasitic growth rates are smaller than
 unstratified ones. If this is taken into account we obtain
 $b_\text{cr}\sim 100$ for both $R=1/10$ and $R=1/100$.
 Consequently, by the time a
parasite can make a successful attack, the channel will have reached
enormous amplitudes, decreasing the midplane plasma beta by a factor
$10^4$. By then it is more than likely that equipartition has been reached.


\begin{thebibliography}{100}


\bibitem{Arfken}
Arfken, G., 1970. Mathematical methods for physicists (2nd ed.).
Academic Press, London.

\bibitem{BH91}
Balbus, S.~A., Hawley, J.~F., 1991. 
ApJ, 376, 214.

\bibitem{BH92}
Balbus, S.~A., Hawley, J.~F., 1992.
ApJ, 400, 610.


\bibitem{BH98}
Balbus, S.~A., Hawley, J.~F., 1998.      
        Rev.~Mod.~Phys., 70, 1.

\bibitem{BM}
Barranco, J.~A., Marcus, P.~S., 2005.
ApJ, 623, 1157.


\bibitem{B37}
Bickley, W.~G., 1937. 
Phil.~Mag., 23, 727.


\bibitem{BSZ98}
Biskamp, D., Schwarz, E., Zeiler, A., 1998.
Phys.~Plasmas, 5, 2485.

\bibitem{Black}
Blackman, E.~G., Pessah, M.~E., 2009.
ApJ, 704, 113.

\bibitem{BMCRF08}
Bodo, G., Mignone, A., Cattaneo, F., Rossi, P., Ferrari, A., 2008.
A\&A, 487, 1.


\bibitem{B00}
Boyd, J.~P., 2000. Chebyshev and Fourier Spectral Methods (2nd ed.).
Dover Publications, New York.

\bibitem{Brand}
Brandenburg, A., Nordlund, A., Stein, R.~F., Torkelsson, U., 1995.
ApJ, 446, 741.

\bibitem{Detal}
Davis, S.~W., Stone, J.~M., Pessah, M.~E.,
2010. ApJ, submitted (arXiv:0909.1570).


\bibitem{DrRe}
Drazin, P.~G., Reid, W.~H., 1982. Hydrodynamic stability.
Cambridge University Press, Cambridge UK.


\bibitem{FlemStone}
Fleming, T., Stone, J.~M., 2003.
ApJ, 585, 908.

\bibitem{Fromangetal}
Fromang, S., Hennebelle, P., Teyssier, R., 2006.
A\&A, 457, 371.

\bibitem{GammieB}
Gammie, C.~F., Balbus, S.~A., 1994.
MNRAS, 270, 138.

\bibitem{GammieM}
Gammie, C.~F., Menou, K., 1998.
ApJL, 492, L75.
\bibitem{GilGlatz}
Gilman, P.~A., Glatzmaier, G.~A., 1981.
ApJS, 45, 335. 

\bibitem{GLB65}
Goldreich, P., Lynden-Bell, D., 1965.
MNRAS, 130, 125.


\bibitem{Golub}
Golub, G.~H., Van Loan, C.~F., 1996.
Matrix Computations (3rd ed.). John Hopkins Uni Press, Baltimore.

\bibitem{GX94}
Goodman, J., Xu, G., 1994. ApJ, 432, 213.

\bibitem{Gof}
Gough, D.~O., 1969. JAtS, 26, 448.

\bibitem{H00}
 Hawley, J.~F., 2000. ApJ, 528, 462.

\bibitem{H01}
Hawley, J.~F., 2001. ApJ, 554, 534.

\bibitem{HGB95}
 Hawley, J.~F., Gammie, C.~F., Balbus, S.~A, 1995. 
        ApJ, 440, 742.

\bibitem{Hirose}
Hirose, S., Krolik, J.~H., Stone, J.~M., 2006.
ApJ, 640, 901.


\bibitem{JohLev}
Johansen A., Levin, Y., 2008.
A\&A, 490, 501.


\bibitem{LLB09}
Latter, H.~N., Lesaffre, P., Balbus, S.~A., 2009.
MNRAS, 394, 715.

\bibitem{Liverts}
Liverts, E., Mond, M., 2009.
MNRAS, 392, 287.

\bibitem{MeshSinai}
Meshalkin, L., Sinai, Y., 1961.
J.Appl.Math.Mech, 25, 1700.


\bibitem{Mil&Stone}
Miller, K.~A., Stone, J.~M., 2000.
ApJ, 534, 398.

\bibitem{Ogil1}
Ogilvie, G.~I., 1998.
MNRAS, 297, 291.


\bibitem{Ogil2}
Ogilvie, G.~I., Livio, M., 2001.
ApJ, 553, 158.

\bibitem{Papa}
Papaloizou, J., Szuszkiewicz, E., 1992.
GAFD, 66, 223.


\bibitem{Pess}
Pessah, M.~E., 2009.
eprint arXiv:0908.1791

\bibitem{Pes&Good}
Pessah, M.~E., Goodman, J., 2009.
ApJ, 698, 72.

\bibitem{Salmon1}
Salmeron, R., Wardle, M., 2003.
MNRAS, 345, 451.

\bibitem{Salmon2}
Salmeron, R., Wardle, M., 2005.
MNRAS, 361, 45.

\bibitem{Salmon3}
Salmeron, R., Wardle, M., 2008.
MNRAS, 388, 1223.

\bibitem{Salmon4}
Salmeron, R., K\"onigl, A., Wardle, M., 2007.
MNRAS, 375, 177.

\bibitem{Sano}
Sano, T., 2007. Ap\&SS, 307, 191. 

\bibitem{SM99}
Sano, T., Miyama, S.~M., 1999. 
ApJ, 515, 776.

\bibitem{SanoInu}
Sano, T., Inutsuka, S., 2001.
ApJ, 561, 179.

\bibitem{Shi}
Shi, J., Krolik, J.~H., Hirose, S., 2010.
ApJ, 708, 1716.

\bibitem{Stoneetal}
Stone, J.~M., Hawley J.~F., Gammie C.~F., Balbus S.~A.,
1996. ApJ, 463, 656.

\bibitem{SuzInu2}
Suzuki, T.~K., Inutsuka, S., 2009.
ApJ, 691, 49.

\bibitem{Teyssier02}
Teyssier, R., 2002.
A\&A, 385, 337.

\bibitem{Zhu}
Zhu, Z., Hartmann, L., Gammie, C., McKinney, J.~C., 2009.
ApJ, 701, 620.

\bibitem{Ziegler}
Ziegler, U., 2004.
JCoPh, 196, 393.





\end{thebibliography}
\end{document}